\documentclass[11pt,letterpaper]{article}
\pdfoutput=1
\usepackage{jheppub}
\usepackage{bbm}
\usepackage{mathrsfs}
\usepackage{slashed}
\usepackage{caption}
\usepackage{epstopdf}
\usepackage[normalem]{ulem}
\usepackage[bottom]{footmisc}
\usepackage{subcaption}
\usepackage{bbold}
\usepackage{titlesec}
\usepackage{threeparttable}
\usepackage{booktabs}
\usepackage{changepage}
\usepackage[utf8]{inputenc}
\usepackage{dsfont}
\usepackage{comment}
\usepackage[dvipsnames]{xcolor}

\usepackage{grffile}
 
\usepackage{graphicx}  % needed for figures
\usepackage{dcolumn}   % needed for some tables
\usepackage{bm}        % for math
\usepackage{amssymb}   % for math
\usepackage{setspace}
\usepackage{amsmath, amssymb, setspace}
\usepackage{array}
\usepackage{booktabs}
\usepackage{caption}
\usepackage{indentfirst}
\usepackage{float}
\usepackage{lmodern}
\usepackage{multirow}
\usepackage{soul}
\usepackage[normalem]{ulem}

\def\a{\alpha}  \def\g{\gamma} \def\e{\epsilon}
\def\d{\delta}
\def\g{\gamma} \def\m{\mu} \def\n{\nu}  \def\r{\rho}
 \def\o{\omega}  \def\l{\lambda}
\def\D{\Delta} \def\G{\Gamma} \def\O{\Omega} \def\L{\Lambda}

 \newcommand{\Bcal}{{\mathcal B}}
\newcommand{\Ccal}{{\mathcal C}} 
 \newcommand{\Fcal}{{\mathcal F}}

\newcommand{\Lcal}{{\mathcal L}}
 
\newcommand{\Ocal}{{\mathcal O}}
 \newcommand{\Scal}{{\mathcal S}}

\newcommand{\nn}{\nonumber}
\def\ddx#1#2{\frac{d #1}{d #2}}

\newcommand{\Mpl}{M_{\rm Pl}}
 
\title{\huge{Diffuse Axion Background}}

\author[a,b]{Joshua Eby}
\author[c,d,e,b]{and Volodymyr Takhistov} 

\affiliation[a]{The Oskar Klein Centre, Department of Physics, Stockholm University, 10691 Stockholm, Sweden}
\affiliation[b]{Kavli Institute for the Physics and Mathematics of the Universe (WPI), Chiba 277-8583, Japan}
\affiliation[c]{International Center for Quantum-field Measurement Systems for Studies of the Universe and Particles (QUP,WPI), High Energy Accelerator Research Organization (KEK), Oho 1-1, Tsukuba, Ibaraki 305-0801, Japan}
\affiliation[d]{Theory Center, Institute of Particle and Nuclear Studies (IPNS),
High Energy Accelerator Research Organization (KEK), Tsukuba 305-0801, Japan}
\affiliation[e]{Graduate University for Advanced Studies (SOKENDAI), \\
1-1 Oho, Tsukuba, Ibaraki 305-0801, Japan}

\emailAdd{joshua.eby@fysik.su.se} 
\emailAdd{vtakhist@post.kek.jp} 

\abstract{ 
Relativistic axions can be readily produced in a broad variety of transient sources, such as axion star bosenova explosions, supernovae or even evaporating primordial black holes. We develop a general framework describing the resulting persistent diffuse axion background (D$a$B) due to accumulated axions
from historic transient events. We derive strong constraints on the D$a$B flux from light axions  $m\lesssim 10^{-3}\,{\rm eV}$ emitted from sources with energies $\o\gtrsim{\rm MeV}$ considering the non-observation of excess photons associated with axion-photon coupling from experiments, including COMPTEL, NuSTAR, XMM-Newton, INTEGRAL, EGRET and Fermi. Future searches in experiments such as SKA, JWST, XRISM, Vera C. Rubin Observatory, AMEGO/e-ASTROGAM will allow probing D$a$B and associated axion-photon couplings with unprecedented sensitivity covering a wide range of possible source energies as low as $0.1\,\mu$eV and multiple decades in axion masses. We highlight the differences between astrophysical and dark sector sources of D$a$B. Further, we discuss complementarity with direct detection as well as prospects for other D$a$B searches. Our analysis demonstrates that D$a$B can act as a promising probe of populations of axion emission sources as well as emission mechanisms.}

\begin{document}
\preprint{KEK-QUP-2024-0001, KEK-TH-2596, KEK-Cosmo-0337, IPMU24-0003}
 \maketitle
\flushbottom
 
\section{Introduction}

The discovery of Higgs boson~\cite{ATLAS:2012yve,CMS:2012qbp} has highlighted the central role of scalar spin-0 particles in Nature. Existence of pseudo-scalar spin-0 quantum chromodynamics (QCD) axions would readily address 
the long-standing strong CP problem of the Standard Model (SM)~\cite{Peccei:1977hh,Weinberg:1977ma,Wilczek:1977pj,Kim:1979if,Shifman:1979if,Zhitnitsky:1980tq,Dine:1981rt} and provide a compelling candidate for the dark matter (DM)~\cite{Preskill:1982cy,Abbott:1982af,Dine:1982ah}. Axions and more generally axion-like particles\footnote{Here, we denote axion and axion-like particles as \emph{axions}.}, pseudo-scalars that generically arise from broken shift symmetries, are expected to be ubiquitous from fundamental theory~\cite{Svrcek:2006yi,Arvanitaki:2009fg}.
Significant efforts have been devoted to explore the axion landscape~(see e.g.~\cite{Irastorza:2018dyq,Agrawal:2021dbo,Adams:2022pbo} for an overview), primarily focusing on non-relativistic cold axion DM.

Relativistic axions can appear from a variety of sources. Cosmological axion production at redshifts prior to galaxy and star formation ($z \gtrsim 30$) will result in new contributions to the cosmic axion background~\cite{Conlon:2013isa,Marsh:2013opc,Dror:2021nyr}. Another prominent possibility are thermal axions, which are produced in the early Universe if axions were in thermal equilibrium through interactions with SM constituents~\cite{Turner:1986tb,Chang:1993gm,Masso:2002np,Hannestad:2005df,Graf:2010tv,Salvio:2013iaa,Arias-Aragon:2020shv,Dror:2021nyr}. Efficient non-thermal cosmological axion production is also possible, such as due to particle decays~\cite{Chun:1995hc,Asaka:1998xa,Ichikawa:2007jv,Kawasaki:2007mk} or non-linear dynamics of field oscillations~\cite{Kasuya:1996ns,Ema:2017krp,Co:2017mop}.
On the other hand, continuous emission at low-redshifts of relativistic axions from astrophysical sources like stars and the Sun has resulted in variety of sensitive tests of possible axion parameter space, such as by considerations of star cooling~\cite{Raffelt:1996wa} or direct detection searches for solar axion emission~\cite{Sikivie:1983ip,Moriyama:1995bz}.

Recently, a fruitful new direction focusing on manifestations of relativistic burst axion emissions from transient sources has emerged and is being explored, including axion (boson) star explosions~\cite{Eby:2016cnq,Levkov:2016rkk}, black hole superradiance~\cite{Arvanitaki:2010sy,Arvanitaki:2014wva,Arvanitaki:2016qwi,Baryakhtar:2020gao,Unal:2020jiy,Branco:2023frw}, and solar halos~\cite{Banerjee:2019epw,Banerjee:2019xuy,Budker:2023sex} that can repeatedly explode on astrophysical timescales.
Direct detection of relativistic bosons emitted from transient sources was found to result in a rich variety of signatures in the context of axion searches~\cite{Eby:2021ece,Chigusa:2023bga} as well as precision quantum sensors~\cite{Arakawa:2023gyq}, often allowing to probe novel parameter space that is beyond the reach of conventional cold DM searches. Furthermore, such axion manifestations hold promising avenues for exploration of new physics
in the context of multimessenger astronomy~\cite{Dailey:2020sxa,Eby:2021ece, Arakawa:2023gyq}.

Accumulation of relativistic axions from historic transient sources will result in persistent axion flux filling the Universe~\cite{Raffelt:2011ft,Calore:2020tjw,Calore:2021hhn,Takhistov:2021qzv}. We comprehensively explore this~\textit{diffuse axion background} (D$a$B). An analogy can be drawn between D$a$B and neutrinos emitted from historic astrophysical sources, such as supernovae, which will contribute to a diffuse supernova neutrino background~(see e.g.~\cite{Beacom:2010kk} for review), detection of which is a target of ongoing efforts by Super-Kamiokande~\cite{Horiuchi:2008jz,Super-Kamiokande:2021jaq} and other neutrino experiments, or other possible sources such as supermassive star explosions~\cite{Shi:1998nd,Linke:2001mq,Munoz:2021sad}. Proposed searches for axions contributing to cosmic axion background, e.g.~\cite{Conlon:2013txa,Angus:2013sua,Jaeckel:2021ert,Cui:2022owf,Kar:2022ngx,ADMX:2023rsk}, could also be in principle considered for diffuse axion background.

In this work, we develop a general framework and systematically analyze D$a$B originating from distinct sources. Depending on the origin, D$a$B contributions can dramatically differ. We discuss variety of general approaches for D$a$B detection, including with existing as well as future experiments. 

The format of this paper is as follows. In Sec.~\ref{sec:dabflux} we introduce the general framework for D$a$B and discuss different possible contributing sources. In Sec.~\ref{sec:axionphoton} we focus on photon production from D$a$B. Then, in Sec.~\ref{sec:signals} we analyze observable signatures with photons originating from D$a$B for different existing as well as future experiments. Later, in
Sec.~\ref{sec:astarbosenova} we focus on analyzing D$a$B associated with axion star bosenovae emission as a characteristic source example. We discuss other D$a$B search prospects in Sec.~\ref{sec:additional}, including different axion couplings as well as direct detection. We conclude in Sec.~\ref{sec:conclusions}.

\section{Diffuse axion background flux}
\label{sec:dabflux}

\begin{figure} 
\begin{center}
  \includegraphics[width=1\textwidth]{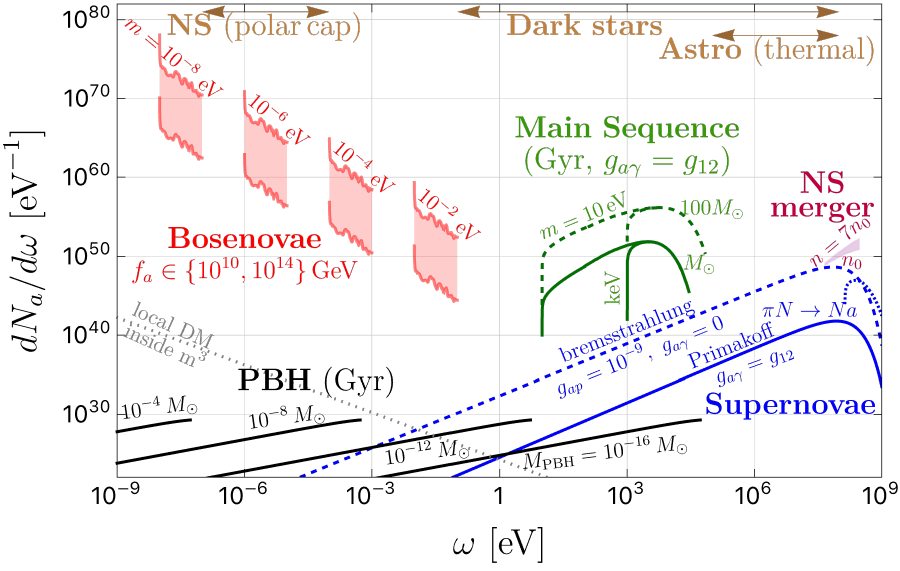}  
	\caption{Sources of relativistic axion flux emission, which can contribute to the D$a$B as described in Sec.~\ref{ssec:sources}. {\color{blue}\bf Blue:} Supernova (SN) emission estimated from Ref.~\cite{Calore:2020tjw} (blue dashed corresponding to brehsstrahlung with axion proton coupling $g_{a p} = 10^{-9}$, $g_{a \gamma} = 0$ allowed by constraints from excessive SN cooling  ~\cite{Carenza:2019pxu}, axion-photon coupling, blue solid for Primakoff processes with $g_{a \gamma} = g_{12} \equiv 10^{-12}$ GeV$^{-1}$, and $\pi N\to N a$ production using $g_{a p} = 10^{-9}$~\cite{Carenza:2020cis,Lella:2022uwi}). 
    {\color{red}\bf Red:} Axion star bosenova emission estimated from~numerical simulations of Ref.~\cite{Levkov:2016rkk} (shaded region corresponds to axion decay constant between $f_a = 10^{14}$~GeV, top, and $f_a = 10^{10}$~GeV, bottom). 
    {\color{black}\bf Black:} Primordial black hole (PBH) emission through Hawking radiation (emission timescale $\Delta t={\rm Gyr}$, PBH mass $M_{\rm PBH}\in\{10^{-4},10^{-8},10^{-12},10^{-16}\}M_\odot$, assuming $m\ll \o$).
    {\color{OliveGreen}\bf Green:} Stellar emission from Main Sequence stars~\cite{Nguyen:2023czp} (solid and dashed lines correspond to stellar masses of $M_\odot$ and $100M_\odot$, respectively, fixing $\Delta t={\rm Gyr}$ and $g_{a\g}=g_{12}$). 
    {\color{Purple}\bf Purple:} Neutron star (NS) merger emission estimated from Ref.~\cite{Harris:2020qim,Fiorillo:2022piv} (with the energy fixed by the temperature as $\o/3 = T = (20-100)\,{\rm MeV}$, density between $n=(1-7)n_0$, and neutron coupling fixed to $g_{an}=1.5\times 10^{-9}$). 
 {\color{brown}\bf Brown:} Expected energy range for NS polar cap emission, astrophysical thermal emission, and dark star thermal emission are denoted by arrows.
 Reference flux from the local DM density inside an experimental volume of ${\rm m}^3$ is given by the gray dotted line.}
	\label{fig:fluxsources}
\end{center}
	\vspace{-4mm}
\end{figure}

The present-day accumulated D$a$B number flux spectrum that originates from historic transient sources can be stated in generality in analogy with the well-studied diffuse background flux of SM weakly interacting neutrinos originating from sources like historic core-collapse supernovae~\cite{Beacom:2010kk,Ando:2023fcc} (or e.g. supermassive star explosions~\cite{Shi:1998nd,Shi:1998jx,Munoz:2021sad}) as\footnote{The conventional units for emission spectrum $\phi$ are cm$^{-2}$s$^{-1}$.}
\begin{equation} \label{eq:dphido}
    \ddx{\phi}{\omega} = \int_0^\infty dz 
        (1+z)\ddx{N_a(\omega(1+z))}{\omega} R_{\rm burst}(z)
                \left|\frac{dt}{dz}\right|~,
\end{equation}
where $\omega$ is the axion energy, $R_{\rm burst}(z)$ is the rate density of transient ``burst'' events emitting axions due to a particular source (in units of events per time and per volume) as a function of cosmological redshift $z$, and $dN_a/d\o$ is the differential spectrum of the number of axions $N_a$ per unit energy $\o$ characterizing emitted axions from a single transient burst with the $(1+z)$ factor accounting for the energy redshifting of axions emitted at $z>0$.
The last term encodes the cosmology of flat expanding Universe dominated by matter and cosmological constant $\left|dt/dz\right|^{-1} = H_0(1+z)\sqrt{\O_\L + \O_M(1+z)^3}$,
with $H_0 = 67.4$ km s$^{-1}$ Mpc$^{-1}$, and densities $\O_\L = 0.685$ for cosmological constant and $\O_M = 0.315$ for matter compatible with latest measurements from Planck~\cite{Planck:2018vyg}. The flux for various possible sources, including both astrophysical as well as dark-sector sources, is illustrated in Fig.~\ref{fig:fluxsources}. 

Looking ahead, relevant for D$a$B detection are axion interactions with SM, which generally depend on specific theory. More so, for astrophysical transient sources, axion-SM interactions are also generally responsible for their production as energy is converted from SM constituents to axions. A canonical coupling of pseudoscalar axion $a$ is the two-photon coupling
described by the Lagrangian contribution
\begin{equation} \label{eq:Lphigg}
    \Lcal \supset \frac{1}{4} g_{a\g}a F_{\m\n}\tilde{F}^{\m\n} = g_{a\g} a \textbf{E$\cdot$B}~,
\end{equation}
where $F_{\m\n}$ is the electromagnetic field-strength tensor, $\tilde{F}^{\m\n}$ is its dual, and $g_{a\g}$ the axion-photon coupling\footnote{This coupling has typical dimension of GeV$^{-1}$.}. Interacting with external magnetic field \textbf{B}, the coupling of Eq.~\eqref{eq:Lphigg} results in axion-photon conversion that is a prominent source of phenomenological signatures.
Additional possible axion interactions include 
\begin{equation} \label{eq:othercoup}
    \Lcal \supset \frac{g_{a\chi}}{2m_\chi}\partial_\mu a
                \overline{\chi} \gamma^{\mu} \gamma_5 \chi
            + \frac{g_{ag}}{4}a G_{\mu\nu}\tilde{G}^{\mu\nu}\,,
\end{equation}
where $G_{\mu\nu}$ is the gluon field strength tensor, $\tilde{G}_{\mu\nu}$ is its dual, $\chi$  denotes fermions such as nucleons $N$ (protons, neutrons) or electrons $e$, $m_\chi$ is their mass, and $g_{a\chi}$ and $g_{ag}$ are coupling constants. For QCD axion that resolves the CP problem in the SM QCD sector, the axion mass $m$ and couplings are related to the Peccei-Quinn symmetry breaking scale, also called the axion decay constant, $f_a$. In particular, $(m/1~{\rm eV}) = (7\times10^6\,{\rm GeV}/f_a) = 0.5 \xi g_{a\gamma}/(10^{-10}~{\rm GeV}^{-1})$ with $\xi$ being a model-dependent constant, which in KSVZ QCD axion scenario~\cite{Kim:1979if,Shifman:1979if} is $|\xi| \simeq 0.5$ and in DFSZ QCD axion scenario~\cite{Dine:1981rt,Zhitnitsky:1980tq} is $|\xi| \simeq 1.4$. 

\subsection{Cosmological transient event rate}

To capture a broad class of possible D$a$B sources, it is convenient to parameterize the cosmological transient event rate as
\begin{equation} \label{eq:Rnova}
 R_{\rm burst}(z) = \frac{A \rho_{\rm loss} H_0}{E_{
 \rm tot}(z)} f(z)~,
\end{equation}
where $f(z)$ is a dimensionless function of redshift $z$ characterizing transient source distribution. 
Here, the normalization constant $A$ is fixed by the requirement that the energy density $\rho_{\rm loss}$ in axions emitted from a particular transient source type, such as exploding axion stars, constitutes a specified subdominant fraction of the total DM density\footnote{In case of exploding supermassive stars, analogous normalization can be imposed by considering relation with baryon abundance~\cite{Munoz:2021sad}.}
\begin{equation} \label{eq:rholoss}
    \rho_{\rm loss} = \int dt E_{\rm tot}(z) \frac{R_{\rm burst}(z)}{(1+z)^3} 
            = \frac{1}{H_0}\int dz \frac{E_{
 \rm tot}(z) R_{\rm burst}(z)}{(1+z)^{4}\sqrt{\O_\L + \O_M(1+z)^3}}~,
\end{equation}
where $E_{\rm tot}(z)$ is the total energy emitted in each transient event burst at redshift $z$.  Since for the sources we consider the total emitted energy for each event is predominantly independent of redshift, we take $E_{\rm tot}$ to be constant in what follows.
We can set the normalization constant $A$ by inverting Eq.~\eqref{eq:rholoss} as 
\begin{equation} \label{eq:Anorm}
    A = \left[\int dz' \frac{f(z')}{(1+z)^{4}\sqrt{\O_\L + \O_M(1+z)^3}}\right]^{-1}\,,
\end{equation}
which can be directly computed for a given source model $f(z)$. In general, $f(z)$ could have non-vanishing contributions over the redshift range $z\in\{z_{\rm min},z_{\rm max}\}$, where $z_{\rm max}$ is the redshift when explosive transient axion emission events begin to occur and $z_{\rm min}$ when the transient events subside. We consider $z_{\rm min} \rightarrow 0$, although theories of transient axion sources with non-negligible $z_{\rm min}$ can be readily incorporated in our analysis.

The energy density $\rho_{\rm loss}$ can be intuitively expressed as the fraction $\Fcal$ of the average DM energy density in the Universe, $\rho_{\rm DM} \simeq 1.2 \times 10^{-6}\,{\rm GeV/cm}^3$,
as $\rho_{\rm loss} = \mathcal{F} \rho_{\rm DM}$. We consider $\Fcal=0.1$ as a reference benchmark, which can evade various constraints. As we will show below, some of our signatures are detectable even for significantly smaller DM fractions. 

\begin{figure} 
\begin{center}
  \includegraphics[width=0.8\textwidth]{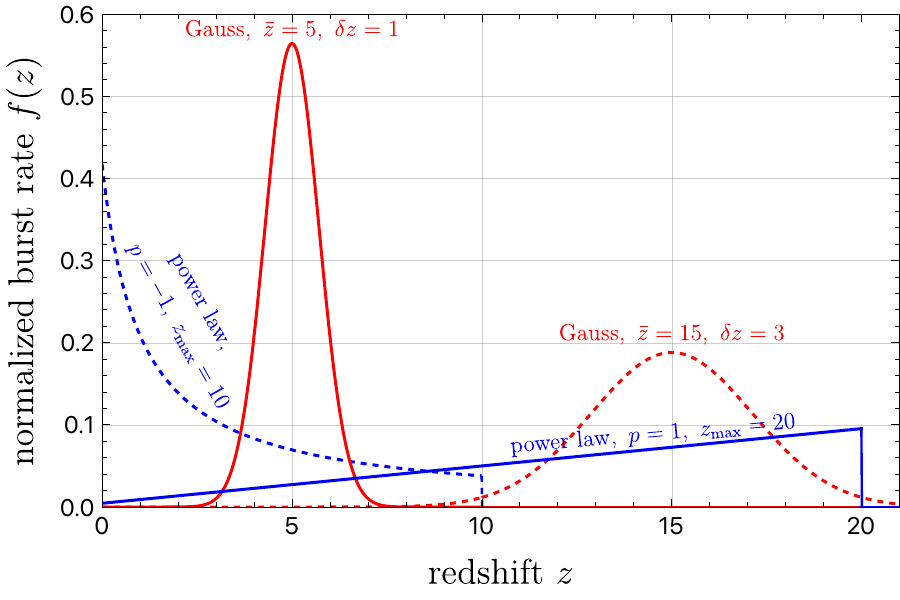}  \hspace{3em}
	\caption{Dimensionless cosmological transient axion burst event rate distribution $f(z)$ sourcing relativistic axions that are considered in this work. The power-law and Gauss distributions are defined in Eq.~\eqref{eq:fzmodels}, and we set the input parameters as labeled. For illustration we normalize $f(z)$ to unity.}
	\label{fig:burstdistributions}
\end{center}
	\vspace{-4mm}
\end{figure}

The cosmological event rate $R_{\rm burst}(z)$ characterizes the epochs during which the transient sources are most active. Employing the normalization of Eq.~\eqref{eq:Rnova}, we analyze two characteristic phenomenological models of source events, described by 
\begin{equation} \label{eq:fzmodels}
    f(z) = \begin{cases}
        (1+z)^p \Theta\left(z-z_{\rm max}\right) & {\rm power\,\,law} \\
        \exp\left[-(z-\bar{z})^2/\d z^2\right]  & {\rm Gauss.}
    \end{cases}\,\,,
\end{equation}
where $p$ is an integer, and $\bar{z}$ with $\d z$ denote the Gaussian distribution mean and width, respectively. The power law is motivated by scenarios where the transient source rate is smoothly varying with possible additional redshift dependence as specified by parameter $p$, reaching a maximum in the present-day for $p<0$ or at earlier times for $p>0$. The Gaussian model instead characterizes general scenarios where transient events predominantly occur during a relatively short cosmological epoch, which could be associated with some dynamical processes, with a duration approximated by $z\in \{\bar{z}-\d z,\bar{z}+\d z\}$. We illustrate several distributions considered in this work in Fig.~\ref{fig:burstdistributions}. Our analysis can readily be carried over to other choices of source event rates, specific to particular theories.

With $f(z)$ given by Eq.~\eqref{eq:fzmodels}, we can directly compute normalization $A$ of Eq.~\eqref{eq:Anorm}.
For the power-law model we find
\begin{equation}
    A_1 = \left[\int_0^{z_{\rm max}} dz' 
        \frac{(1+z')^{p-4}\Theta(z-z_{\rm max})}{\sqrt{\O_\L + \O_M(1+z')^3}}\right]^{-1} 
            \simeq \frac{1}{\sqrt{\O_M}}\left(\frac{9}{2}- p\right) + \Ocal(z_{\rm max}^{p-9/2})~,
\end{equation}
which is valid as long as $p<9/2$, and where we approximated $\sqrt{\O_\L + \O_M(1+z)^3}\simeq \sqrt{\O_M(1+z)^3}$ considering Universe is matter-dominated for the majority of redshift values of interest\footnote{Corrections induced by this approximation are of order $\Ocal(1)$ at most.}. Further corrections are suppressed by powers of $1/z_{\rm max}$, which we consider to be subdominant compared to unity.
For the Gaussian distribution
\begin{equation} \label{eq:anormgauss}
    A_2 = \left[\int_0^{z_{\rm max}}
    dz' \frac{\exp\left(-\left(\frac{z'-\bar{z}}{\delta z}\right)^{2}\right)}{(1+z')^4\sqrt{\O_\L + \O_M(1+z')^3}}\right]^{-1}\,.
\end{equation}
Evaluating Eq.~\eqref{eq:anormgauss} numerically, one finds $A_2\simeq 1-10^{6}$ upon varying $\d z\simeq 10^{-2}-1$ and $\bar{z}\simeq 0-20$. In this estimate we used $z_{\rm max}=20$, but the results do not depend significantly on this choice as long as $z_{\rm max}>\bar{z}$ is satisfied. In what follows, we focus on the power-law model for simplicity and outline results for the Gaussian model in App.~\,\ref{app:RateFunc}.

\subsection{Emission sources}
\label{ssec:sources}

The flux of axions in the relativistic background, given in Eq.~\eqref{eq:dphido}, depends sensitively on the spectrum of emission from transient sources $dN_a/d\o$. 
Our general analysis framework can be applied to a wide range of possible transient sources. A generic description of a peaked axion burst can be captured by a Gaussian function
\begin{equation} \label{eq:dNado_Gaus}
    \frac{dN_a(\o)}{d\omega}\Bigg|_{\rm Gauss} 
        = C \frac{E_{\rm tot}}{m^2} 
            \frac{\exp\left(-\left(\frac{\omega-\bar{\o}}{\d\o}\right)^2\right)}{\sqrt{\pi}\,(\d\o/m)}\,,
\end{equation}
with a central peak energy $\bar{\o}$ and a width $\d\o$. 
The normalization of the emission spectrum $m\int d\o (dN_a/d\o) = E_{\rm tot}$ implies that the normalization constant is given by 
\begin{align} \label{eq:Cnorm}
    C = m\left[\int \frac{\exp\left(-\left(\frac{\omega-\bar{\o}}{\d\o}\right)^2\right)}{\sqrt{\pi}\,(\d\o/m)} d\omega\right]^{-1}
      &= 2\left[1+{\rm erf}\left(\frac{\bar{\o}}{\d\o}\right)\right]^{-1}\,.
\end{align}
Note that $C$ is independent of $m$ and $E_{\rm tot}$. 

We can employ the Gaussian function description to determine the axion background density for a generalized emission source, considering the emission follows an approximately peaked structure. 
In the limit $\d\o\to 0$, $(dN_a/d\omega)_{\rm Gauss} \to (C E_{\rm tot}/m) \delta(\omega-\bar{\o})$, capturing the effect of a narrow burst spectrum in the $\d\o\ll m$ limit.
Combining Eqs.~\eqref{eq:dphido}, \eqref{eq:Rnova}, \eqref{eq:dNado_Gaus}, and \eqref{eq:Cnorm}, we obtain
\begin{equation} \label{eq:dphido_Gaus}
    \ddx{\phi}{\omega} \Big|_{\rm Gauss} = \frac{\Fcal \r_{\rm DM}}{m \d\o}\frac{2}{\sqrt{\pi}}\frac{A}{\left[1+{\rm erf}(\bar{\o}/\d\o)\right]}
        \int \frac{dz f(z)}{\sqrt{\O_\L + \O_M(1+z)^3}}
        \exp\left[-\left(\frac{(\o(1+z)-\bar{\o})}{\d\o}\right)^2\right]\,.
\end{equation}
We stress that the flux model of Eq.~\eqref{eq:dphido_Gaus} captures key behavior while being particularly straightforward to analyze. For a given particle mass $m$ and cosmological rate function $f(z)$, one needs only to specify the central burst energy $\bar{\o}$, the emission width $\d\o$, and the total DM fraction converted to relativistic particles $\Fcal$. 

We systematically discuss below several prominent examples of transient axion burst sources resulting in D$a$B 
highlighting the generality and reach of our analysis. 
In Fig.~\ref{fig:fluxsources} we summarize axion emission flux from several characteristic sources, as compared to reference local DM flux within a volume of m$^3$ (see Eq.~\eqref{eq:phiscale_DM}). Analogously to astrophysical multimessenger targets, astrophysical sources such as supernovae can serve as prime production sites of relativistic axions\footnote{We note that some astrophysical axion emission mechanisms can persist over significant timescales. Here, we comment on them as characteristic and reference sources in the context of potentially contributing to D$a$B.}.
This exemplifies conversion of energy associated with SM constituents (e.g. baryons) into dark sectors with large kinetic energy. In this work, in addition to astrophysical sources, we emphasize the significant potential of dark-sector dynamics to produce relativistic axions resulting in sizable contributions to D$a$B, which we can describe using a general and flexible framework. 
We further illustrate how this can give rise to photon signatures across a broad variety of source classes, range of energy scales, and distinct photon-production scenarios.

\subsubsection{Supernovae}
\label{ssec:Supernova}

Astrophysical sources could serve as powerful production sites not only of neutrinos, but also of other weakly interacting particles such as axions. Core-collapse supernovae (SN) allow to sensitively probe axions through variety of methods~(e.g.~\cite{Raffelt:1990yz,Raffelt:1996wa,Raffelt:2006cw,Fischer:2016cyd,Caputo:2024oqc}). Aside first direct SN neutrino observations~\cite{Kamiokande-II:1987idp,Bionta:1987qt,Alekseev:1988gp}, SN1987A  that exploded in the Large Magellanic Cloud at
a distance of 50 kpc proved to be an essential laboratory for axion physics~(e.g.~\cite{Turner:1987by,Burrows:1988ah,Burrows:1990pk,Raffelt:1987yt,Raffelt:1990yz,Keil:1996ju,Hanhart:2000ae,Chang:2018rso}). In SN, efficient axion production can occur through axion nucleon-nucleon 
bremsstrahlung process~\cite{Turner:1987by} $N N \rightarrow N N a$, where $N$ denotes nucleons (protons, neutrons). Refs.~\cite{Carenza:2020cis,Lella:2022uwi} have illustrated the importance of a pion production channel $\pi N \rightarrow N a$ as well, which can dominate axion emission above energies of order $\sim 150\,{\rm MeV}$. Requiring that SN1987A neutrino emission is not significantly reduced by axion emission, we consider axion-nucleon coupling\footnote{Exact value depends on specific considered axion model.} $g_{a N} \lesssim 10^{-9}$~\cite{Carenza:2019pxu}. When axions couple only to photons with coupling $g_{a\gamma}$, efficient production can occur through Primakoff process when thermal photons are converted into axions within electromagnetic field of stellar medium~\cite{Raffelt:1985nk}.
This results in strong limits on axions from SN energy loss arguments more stringent than other limits above tens of keV~\cite{Jaeckel:2010ni,Caputo:2021rux,Caputo:2022rca,Lella:2023bfb}. For light SN axions constraints arise from conversion to photons 
in Milky Way Galactic magnetic field~\cite{Grifols:1996id,Brockway:1996yr,Payez:2014xsa}. For heavy SN axions constraints arise from decays along the emitted axion line of sight and non-observation of SN1987A coincidence signal~\cite{Jaeckel:2017tud}.

Contributions to D$a$B from SN~\cite{Raffelt:2011ft,Calore:2020tjw,Calore:2021hhn} can be effectively parameterized using a fit to the axion spectrum arising due to a given process channel $x$ of the form~\cite{Calore:2020tjw}
\begin{equation}
    \frac{dN_a(\o)}{d\omega}\Bigg|_{\rm supernova} = c\left(\frac{g_{a x}}{g_{a x}^{\rm ref}}\right)^2
    \left(\frac{\omega}{\omega_0}\right)^\beta \exp\left(-\frac{(\beta+1)\omega}{\omega_0}\right)\,,
\end{equation}
where $c$, $g_{ax}^{\rm ref}$, $\omega_0$, and $\beta$ are fit parameters which depend on the production process (see e.g. Tab.~1 of Ref.~\cite{Calore:2020tjw}). Hypernovae, transient sources similar to supernovae but with higher energy and which give rise to a black hole in the final state, are also efficient sources of axions, with a spectrum that peaks around $dN_a/d\o\sim 10^{44}/{\rm eV}$ per second (for $g_{a\g}\simeq 10^{-12}\,{\rm GeV}^{-1}$) between $10-100$~MeV~\cite{Caputo:2021kcv}.

Importantly, for astrophysical transients such as SN contributing to D$a$B the event rate $R_{\rm burst}$ is determined by conventional astrophysical processes (e.g. star formation, stellar nuclear fusion) such as in case of diffuse neutrino background.
However, this is distinct for D$a$B contributions arising from dark sector transient sources such as axion stars explosions.

\subsubsection{Axion star bosenovae} 
\label{ssec:Bosenova}

Axion stars are gravitationally-bound states formed from ultralight bosons~\cite{Kaup:1968zz,Ruffini:1969qy,BREIT1984329,Colpi:1986ye}. 
They can appear in the early Universe within the in cores of diffuse axion miniclusters~\cite{Kolb:1993zz,Eggemeier:2019jsu}. When their mass reaches a critical value $M_c$~\cite{Chavanis:2011zi,Chavanis:2011zm,Barranco:2010ib,Eby:2014fya,Schiappacasse:2017ham}, defined by the strength of the axion self-interaction coupling, these states undergo gravitational collapse and finally emit a large fraction of their mass-energy in relativistic bosons~\cite{Eby:2016cnq,Levkov:2016rkk}. This explosive production of relativistic particles signifies\footnote{The term was originally coined to describe an explosive process in atomic physics~\cite{Donley_2001}.} \emph{bosenova}.

Detailed numerical simulations of boson star collapse and explosive axion emission were carried out in Ref.~\cite{Levkov:2016rkk}. The spectrum of emitted axions $dE/dk$ has been analyzed up to $k/m\simeq 9$, where $E(\o) = \o N_a(\o)$ is the total energy emitted in axions of energy $\o$, and $k = \sqrt{\o^2 - m^2}$ is the corresponding momentum. The total energy emitted is a linear function of the axion decay constant $f_a$, well-fitted by $E_{\rm tot}\simeq M_c[0.3 + 10^3 f_a/\Mpl]$, which saturates to unity near $f_a\simeq 10^{16}\,{\rm GeV}$, and where $M_c=10\Mpl f_a/m$. To analyze resulting D$a$B flux using Eq.~\eqref{eq:dphido}, we translate
\begin{equation} \label{eq:dNado}
    \ddx{N_a(\o)}{\o}\Bigg|_{\rm bosenova} = \left[\ddx{E}{k} - \ddx{\o}{k}\right] \frac{1}{\o \displaystyle{\ddx{\o}{k}}} 
                \simeq \ddx{E}{k} \frac{1}{\sqrt{\o^2 - m^2}}~,
\end{equation}
where in the last step we used $d\o/dk = k/\o$, and neglected the second term in the bracket, which is suppressed by a factor $1/N_a \ll 1$ relative to the leading term.  
In this case, Eqs.~\eqref{eq:dphido} and \eqref{eq:Rnova} give
\begin{align} \label{eq:dphido_BN}
\ddx{\phi}{\o}\Bigg|_{\rm bosenova} &\simeq 
        \frac{\Fcal \r_{\rm DM} A}{\sqrt{\o^2-m^2}}
            \int_0^\infty dz 
        \frac{f(z)}{\sqrt{\O_\L + \O_M(1+z)^3}}
            \frac{1}{E_{\rm tot}}\ddx{E(\o(1+z))}{k}~.
\end{align}
\begin{figure} 
\begin{center}
  \includegraphics[width=0.85\textwidth]{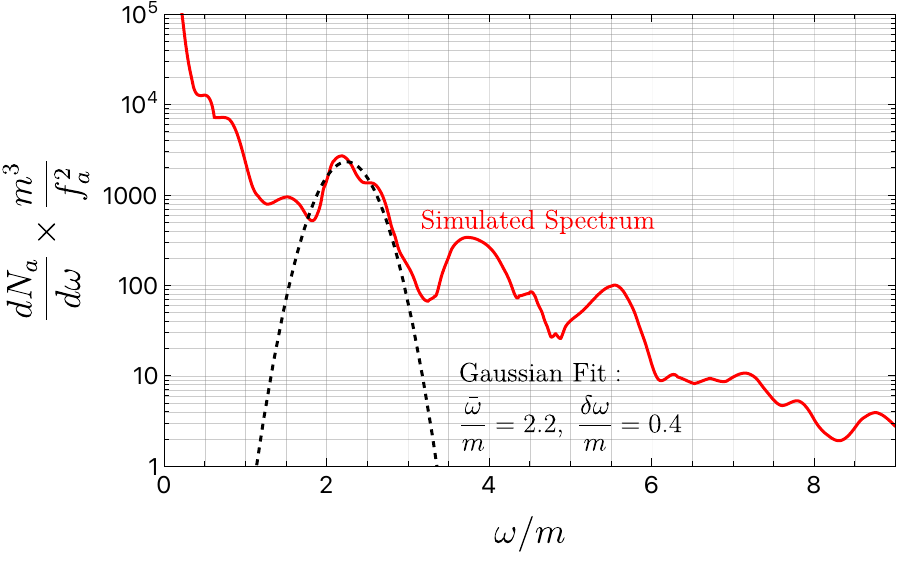}  \hspace{3em}
	\caption{Spectrum of emitted axions as a function of energy, from numerical simulations of axion bosenova explosion~\cite{Levkov:2016rkk} (red solid) with a Gaussian fit (black dashed) using Eq.~\eqref{eq:dNado_Gaus} to the leading relativistic peak at $\bar{\o}\simeq 2.2m$ and $\d\o\simeq 0.4 m$.}
	\label{fig:bosenovaspectrum}
\end{center}
	\vspace{-4mm}
\end{figure}

Key properties of these axion star explosions can be obtained from the emission spectrum~\cite{Levkov:2016rkk}, which is peaked at semi-relativistic energies (i.e. momenta of order $\sim$few$\times m$). The peaks can be understood as a consequence of number-changing processes~(see e.g. Ref.~\cite{Eby:2015hyx,Eby:2017azn}). In Fig.~\ref{fig:bosenovaspectrum} we illustrate the full simulated spectrum of Ref.~\cite{Levkov:2016rkk}, along with a Gaussian fit of the first relativistic peak, which is well-described by Eq.~\eqref{eq:dNado_Gaus} with $\bar{\o}\simeq 2.2 m$ and $\d\o\simeq 0.4$ as shown. Explosive bosenova emissions across astrophysical timescales constitute a prominent contribution to D$a$B arising from dark-sector sources.

Recently, our analyses of direct detection signatures from bosenova explosions occurring inside the Milky Way demonstrated that they constitute particularly promising targets for current and near-future experimental searches~\cite{Eby:2021ece,Arakawa:2023gyq}. 
Ref.~\cite{Du:2023jxh,Escudero:2023vgv} suggested that repeated mergers of DM subhalos can give rise to a large abundance of near-critical boson stars, obtaining a DM fraction as large as $\sim10^{-6}$. These stars would collapse at redshift $z\lesssim 30$, with the resulting relativistic axions contributing to D$a$B. Bosenovae occurring at earlier epochs of redshifts $z\gtrsim 30$ can also contribute, albeit with significantly redshifted axion energies, and can be constrained by cosmological observations~\cite{Fox:2023aat}.

\subsubsection{Primordial black hole evaporation}

Relativistic axions can be readily emitted as Hawking radiation when the black hole temperature exceeds the axion mass from evaporating primordial black holes (PBHs), which can form in the early Universe through variety of mechanisms, providing a novel target for direct searches~\cite{Calza:2021czr,Agashe:2022phd,Jho:2022wxd,Calza:2023rjt,Calza:2023iqa}.
The emission rate of axions produced in Hawking radiation  per unit energy per unit time is given by~\cite{Hawking:1974rv,Page:1976df,MacGibbon:1990zk}
\begin{equation}
    \frac{d^2 N_a}{d \o d t} 
        = \frac{1}{2\pi} \frac{\G(m,\o,M_{\rm PBH})}{e^{\o/T_{\rm H}} - 1}\,,
\end{equation}
where $T_{\rm H}=(8\pi G M_{\rm PBH})^{-1} = 1.06~(10^{13}~{\rm g}/M_{\rm PBH})$~GeV is the Hawking temperature. The greybody factor $\G$ can be computed numerically in general (see e.g.~\cite{Arbey:2019mbc,Arbey:2021mbl}), but can be approximated with the geometric-optics limit for high-energy emission as $\G \simeq 27 G^2 M_{\rm PBH}^2 \o^2$. In that limit, we can write
\begin{align} \label{eq:flux_PBH}
    \frac{d N_a}{d \o}\Bigg|_{\rm PBH} 
    &\simeq \frac{27\Delta t}{2\pi}\frac{(G M_{\rm PBH} \o)^2}{e^{8\pi GM_{\rm PBH}\o}-1} 
    \simeq \frac{27\Delta t}{16\pi^2}G M_{\rm PBH} \o \,,
\end{align}
where $\Delta t$ is the total emission time and in the second step we expanded considering $\o\ll T_{\rm H}$. In addition to contributing to the D$a$B, the emitted axions can be promptly converted to photons. 
In Fig.~\ref{fig:fluxsources} we illustrate the flux of axions from PBH evaporation assuming $\Delta t = $~Gyr and that $m \ll \omega$.
Recently, it was shown that intriguing electromagnetic signatures can also appear from decays of sterile neutrinos originating from evaporating PBHs~\cite{Chen:2023tzd,Chen:2023lnj}.

Note that Eq.~\eqref{eq:flux_PBH} assumes the PBH mass does not change appreciably due to emission, which is a good approximation for $M_{\rm PBH}\gtrsim 10^{-16}M_\odot$. In this limit PBHs act as an approximately continuous source of axion production, though transient signatures are possible during the final stages of PBH evaporation.

\subsubsection{Main-sequence stellar emission}

Stars can continuously emit axions, primarily through Primakoff effect and photon coalescence. The emission peak for such axions is near energies of the order of the star's core temperatures, typically $\bar{\omega}\sim {\rm keV}$ for stellar masses $(0.1-100)M_\odot$.
The luminosity $L_a^*$ of such sources was computed in detail in Ref.~\cite{Nguyen:2023czp} for stars in this mass range, which can be converted to a particle flux using
\begin{equation}
    \frac{dN_a}{d\o}(\o)\Bigg|_{\rm stars}
    =\frac{dL_a^*}{d\o} \frac{\Delta t}{\omega} \,,
\end{equation}
with $\Delta t$ being the total emission time. In Fig.~\ref{fig:fluxsources} we illustrate the flux considering $\Delta t = $~Gyr and Primakoff production, which is the dominant contribution for the majority of the parameter space of interest.
This continuous stellar axion emission can readily contribute to D$a$B, as has been recently studied in Ref.~\cite{Nguyen:2023czp}. Beyond main-sequence stars, 
post-main-sequence and population III stars could also potentially source axions and contribute to D$a$B with sufficient core
temperatures.

\subsubsection{Neutron stars}
\label{ssec:NS}

Neutron stars (NSs) constitute another possible prominent astrophysical source of relativistic axion emission.
Axions can be thermally produced inside NSs through various channels, such as nucleon-nucleon bremsstrahlung $N N \rightarrow N N a$ and  Cooper pair-breaking
processes, cooling NSs\footnote{Cooling of white dwarfs and red giants through axion emission can also result in sensitive constraints ~\cite{Altherr:1993zd,Raffelt:1994ry,Corsico:2001be,MillerBertolami:2014rka}.} (e.g.~\cite{Iwamoto:1984ir,Keller:2012yr,Sedrakian:2015krq}). 
Additional production modes are possible in highly-magnetized NSs, i.e. magnetars~\cite{Bai:2023bbg}.
Such emission can be enhanced during NS mergers, a process which has been studied using numerical~\cite{Dietrich:2019shr} and analytic~\cite{Harris:2020qim,Fiorillo:2022piv} approaches. The typical energy scale of emission is near the temperature of a NS $T\simeq \Ocal(1-100)\,{\rm MeV}$, provided the axions are massless by comparison, i.e. $m\ll T$. Furthermore, for typical nuclear couplings allowed by current constraints (see e.g.~\cite{Adams:2022pbo} for review), the axions easily escape the star and free-stream away. 

The axion emissivity in NS mergers as a function of temperature, $Q_{0}(T)$, has been estimated in relativistic mean field theory~\cite{Harris:2020qim} for an axion-neutron coupling of $g_0=1.5\times 10^{-9}$. For other couplings, the emissivity scales as
\begin{equation}
    Q(T) = \left(\dfrac{g_{an}}{g_0}\right)^2Q_{0}(T)\,.
\end{equation}
Following Ref.~\cite{Harris:2020qim}, we use the temperature $T$ as a proxy for the typical axion emission energy as $\o\simeq 3T$.
Taking an approximate merger duration of $t_{\rm merger}\sim 1\,{\rm s}$ and emitting volume of $V\sim (15\,{\rm km})^3$~\cite{Fiorillo:2022piv}, we estimate
\begin{equation}
    \frac{dN_a}{d\o}\Big|_{\rm NS\,merger} \simeq 
        \frac{V\,t_{\rm merger}}{\o^2} Q(T)\,.
\end{equation}
Here, the density of the Fermi fluid $n$ is varied between the nuclear saturation density, $n_0\simeq 10^{-44}\,{\rm m}^3$, and $7n_0$. This range determines the extent of the gray shaded region in Fig.~\ref{fig:fluxsources}. 
Some of the emitted axions convert to gamma rays in the magnetosphere, leading to detectable, transient photon signals~\cite{Fiorillo:2022piv}. Here, we consider the possibility of such axions accumulating over time and contributing to D$a$B. Other axion emission signatures from neutron star mergers are also possible, for example, in the case of fireball formation~\cite{Dev:2023hax,Diamond:2023cto}.

Recently, it was pointed out that axions can be produced directly via a coupling to photons in the polar-cap region of NS~\cite{Prabhu:2021zve,Noordhuis:2022ljw}. While some become trapped around the NS, for axions with masses $m\lesssim 10^{-7}\,{\rm eV}$ the vast majority escape the NS (as many as $\sim 10^{50}$ axions per second), as the energy scale of emission is fixed by dynamics in the polar cap in the range $10^{-7}\,{\rm eV}\lesssim \omega\lesssim 10^{-4}\,{\rm eV}$~\cite{Noordhuis:2023wid}. 
As we discuss further below, such frequency ranges could be probed with future Square Kilometer Array (SKA) observations.

\subsubsection{Examples of other sources}
\label{ssec:additional}

A broad variety of other sources of relativistic axion emission that could contribute to D$a$B are possible, such as:%
\begin{itemize}
\item {\bf Black hole superradiance}: 
``clouds'' of axions can be generated by sapping the angular momentum and energy from rapidly-rotating black holes~\cite{Press:1972zz,Arvanitaki:2010sy,Arvanitaki:2014wva,Arvanitaki:2016qwi,Baryakhtar:2020gao,Unal:2020jiy} (see e.g.~\cite{Brito:2015oca} for review). Superradiance effects around early-Universe (primordial) black holes have also been considered~\cite{Ferraz:2020zgi,Bernal:2022oha,Branco:2023frw}. Massive axion fields that have quasi-bound states around black holes with $\omega < l \Omega_H$, where $\Omega_H$ is the angular velocity of the black hole horizon\footnote{Here, 
$\Omega_H = \frac{1}{r_S}\Big(\frac{a_*}{1+\sqrt{1-a_*^2}}\Big)$, where $r_S = 2 G M_{\rm BH}$ is the black hole Schwarzschild radius and $a_* = J/G M_{\rm BH}^2$ is the dimensionless spin of black hole with angular momentum $J$.
} and $l$ is the angular momentum around
the black hole spin axis, can experience exponentially unstable growth. 
The timescale for spin-down of the BH is give by~\cite{Baryakhtar:2020gao}
\begin{equation}
    \tau_{\rm spin-down} \simeq 10^7\,{\rm years}\left(\frac{0.02}{r_S m}\right)^5\left(\frac{10^{-12}\,{\rm eV}}{m}\right)\left(\frac{0.9}{a_*}\right)^{3/2}\left(\frac{10^{15}\,{\rm GeV}}{f_a}\right)^2\,.
\end{equation}
The growth continues until the black hole loses most of its angular momentum or the superradiance is quenched by the axion self-interactions, either by exciting axions out of superradiant modes leading to stable steady-state configuration or by inducing gravitational collapse. In the latter case, an explosive bosenova-like event occurs near the black hole with rapid emission of relativistic axions~\cite{Arvanitaki:2010sy}. Axion emission can be modified in case of binary black hole mergers~\cite{Baumann:2018vus}.
We note that the event rates of bosenovae-type events associated with black hole supperradiance sensitively depends on a variety of factors such as black hole population statistics, distributions of black hole masses and spins, as well as particle physics effects like interaction couplings of the axions. 

\item {\bf Stellar axion halos}: axions can be captured directly in the gravitational field of astrophysical bodies, such as stars. 
These configurations are akin to gravitational atoms, having a spherically-symmetric density profile of the form~\cite{Banerjee:2019epw,Banerjee:2019xuy}
\begin{equation}
    \rho(r,t) \simeq \rho_0(t) \exp\left(-\frac{r^2}{R_\star^2}\right)\,,
\end{equation}
with radius $R_\star$ and central density $\rho_0(t)$ which can grow exponentially in magnitude with time~\cite{Budker:2023sex}.
Gravitational collapse of these structures around stars across galaxies can potentially lead to occurrence of yet another class of bosenova-like explosive events. The rate of growth happens on a timescales similar to the known relaxation timescales for ultralight particles\footnote{See also~Ref.~\cite{Levkov:2018kau,Chen:2020cef,Kirkpatrick:2020fwd,Chen:2021oot,Kirkpatrick:2021wwz,Dmitriev:2023ipv,Jain:2023tsr} for discussion of relaxation of ultralight fields through gravity and self-interactions.}~\cite{Budker:2023sex}, which for $2\to 2$ self-interactions takes the form
\begin{equation}
    \tau_{\rm rel} \simeq \frac{64 f_a^4 m^3 v_{\rm lDM}^{2}}{\rho_{\rm lDM}^2}
    \simeq 9\,{\rm Gyr}
    \left(\frac{f_a}{10^8\,{\rm GeV}}\right)^4
    \left(\frac{m}{10^{-14}\,{\rm eV}}\right)^3
    \left(\frac{0.4\,{\rm GeV/cm}^3}{\rho_{\rm lDM}}\right)
    \left(\frac{v_{\rm lDM}}{10^{-3}}\right)\,,
\end{equation}
where $\rho_{\rm lDM}\simeq 0.4\,{\rm GeV/cm}^3$ and $v_{\rm lDM} \simeq 10^{-3}$ are the energy density and velocity dispersion, respectively, of local DM in the solar neighborhood vicinity (see e.g.~\cite{Gelmini:2015zpa} for review). Thus, on astrophysical timescales for sufficiently-strong self interactions, solar halos can collapse and explode multiple times for a given star, potentially converting a significant fraction of axionic DM to relativistic D$a$B constituents in the process.

\item {\bf Dark non-bosonic stars:} Astrophysical objects formed from SM particles (e.g. stars and NS) could have analogues in the dark sector, sometimes called \emph{dark stars}~\cite{Maselli:2017vfi,Kamenetskaia:2022lbf}, composed of fermions rather than bosons. If their constituents are coupled to axions, they will be emitted by dynamical processes over a broad range of energies dictated by the dark star core temperature $T_{\rm dark}$, which can be drastically distinct from $\sim {\rm MeV}$ energies associated with ordinary astrophysical stars.
An intriguing possibility is \emph{mirror star}, formed from dark baryons in the context of complex dark sector models~\cite{Curtin:2019lhm,Hippert:2021fch,Hippert:2022huw}. The constituents of dark stars and the interacting SM particles could contribute to axion emission, and therefore to the D$a$B. Population statistics and dynamics of such objects could be dramatically distinct from that of conventional astrophysical stars.

\item {\bf Secondary astrophysical axion production:}
Secondary axions could be produced from decays of dark sector fields originating from various astrophysical processes, such as $\chi \rightarrow aa$ for a dark scalar $\chi$. 
We note that this is distinct from cosmological DM decaying to axions~(e.g.~\cite{Kar:2022ngx}).
    
\end{itemize}

Many relativistic axion emission mechanisms remain poorly understood without a well studied axion emission flux. 
Our analysis highlights the necessity and serves as an impetus for further detailed numerical studies and simulations required to comprehensively describe and quantify the axion emission mechanisms from distinct sources. 
A population of astrophysical sources that produces energetic axions can contribute to D$a$B.

\subsection{Comparison to local DM flux}
\label{sec:DMflux}

A simple estimate of D$a$B flux can be obtained as follows. Assuming a total density $\rho_{\rm loss}$ is converted into relativistic axions of average energy $\bar{\o}$, ignoring redshift effects, the differential flux can be approximated as 
\begin{equation} \label{eq:phiscale}
    \frac{d\phi}{d\o} \sim \frac{\rho_{\rm loss}}{\bar{\o}^2}\,v_{\rm burst}~.
\end{equation} 
This scaling can be roughly obtained from Eq.~\eqref{eq:dphido_BN} in the approximation $dE/dk \to E_{\rm tot}/\bar{\o}$, $\sqrt{\o^2-m^2}\simeq \bar{\o}$ and simultaneously setting $v_{\rm burst} \to 1$, and ignoring the remaining 
prefactor. Similarly we can obtain this scaling from Eq.~\eqref{eq:dphido_Gaus} in the limit\footnote{Here we use the identity \textbf{$\sqrt{2\pi}\d(x)={\rm lim}_{\e\to 0+}\exp(-x^2/2\e^2)/\e$}.}~$\d\o\to 0$, $\o\to\bar{\o}$.

Using Eq.~\eqref{eq:phiscale}, we can estimate a typical ratio of axion background flux to that expected of cold axion DM. Given a local DM density of $\rho_{\rm lDM}\simeq0.4$ GeV/cm$^3$ and velocity dispersion $v_{\rm lDM}\simeq 10^{-3}$ (see e.g.~\cite{Gelmini:2015zpa} for review), the local DM flux is
\begin{equation} \label{eq:phiscale_DM}
    \left(\frac{d\phi}{d\o}\right)_{\rm lDM} \sim \frac{\rho_{\rm lDM}}{m^2}\,v_{\rm lDM} \simeq 10^{26}\,{\rm eV}^{-1}{\rm cm}^{-2}\,{\rm sec}^{-1}\left(\frac{v_{\rm lDM}}{10^{-3}}\right)\left(\frac{10^{-10}\,{\rm eV}}{m}\right)^2~.
\end{equation}
Comparing Eqs.~(\ref{eq:phiscale}-\ref{eq:phiscale_DM}), the expected ratio of D$a$B flux to DM flux to be
\begin{equation} \label{eq:fluxfrac}
    \frac{\displaystyle{d\phi/d\o}}{\displaystyle{\left(d\phi/d\o\right)_{\rm lDM}}}
        \sim \left(\frac{v_{\rm burst}}{v_{\rm lDM}}\right)\left(\frac{m}{\bar{\o}}\right)^2
            \left(\frac{\rho_{\rm loss}}{\rho_{\rm loc}}\right) 
        \simeq  3\cdot10^{-3}\Fcal \left(\frac{m}{\bar{\o}}\right)^2 \,.
\end{equation}
Compared to the local cold DM flux, the D$a$B flux is boosted by its relativistic velocity, $v_{\rm burst}/v_{\rm lDM}\sim 10^3$. However, it is suppressed by the small average DM density of the Universe relative to the expected local DM density, and by $\Fcal<1$, as $\rho_{\rm loss}/\rho_{\rm loc} \sim 10^{-6}\Fcal$.
Finally, for a fixed total energy output the D$a$B is suppressed by the larger energy per particle, $\o$, as $(m/\bar{\o})^2\ll 1$. 
Overall, the D$a$B flux is expected to be at least a few orders of magnitude below that of local cold axion DM in the solar neighborhood.  We stress that these naive estimates do not account for redshift dependencies. Since the energy densities are larger at earlier times, converting a fixed axion DM fraction $\Fcal$ one could convert more axion DM by populations of transient sources at higher redshifts. Hence, Eq.~\eqref{eq:fluxfrac} should be modified by $(1+\bar{z})^3$ scaling when the source distribution is sharply peaked at some redshift $\bar{z}$ in order to properly capture such behavior.

\begin{figure} 
\begin{center}
\includegraphics[width=1\textwidth]{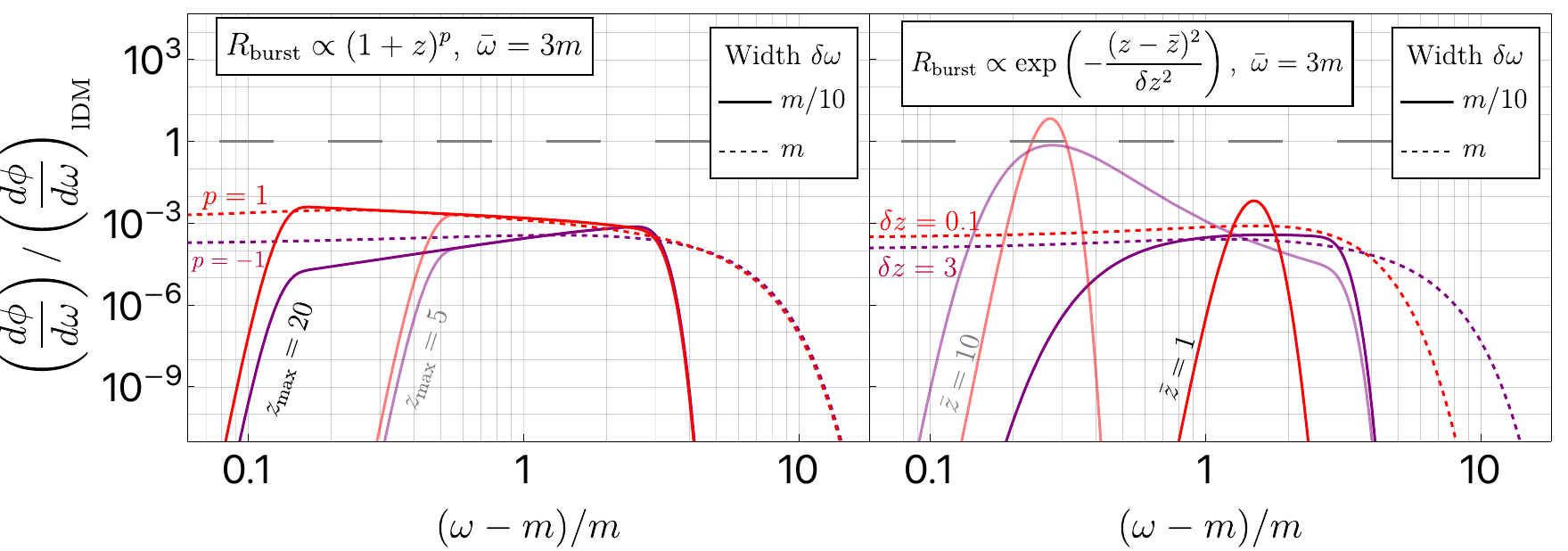} 
	\caption{
 The differential flux,  $d\phi/d\o$, of D$a$B relative to the flux of cold axion DM in the local neighborhood, $(d\phi/d\o)_{\rm lDM}$, as a function of $\o/m$. For the source of the axion background, the thick (dashed) lines correspond to a Gaussian energy distribution at the source with peak energy $\bar{\o}=3m$ and width $\d\o=m$ ($m/10$). The cosmological rate function $R_{\rm burst}$ is given by: (Left panel) a power law $\propto (1+z)^p$ with $p=-1,1$ and $z_{\rm max}=5,20$ as illustrated; (Right panel) a Gaussian $\propto \exp\left[-(z-\bar{z})^2/\d z^2\right]$ with $\bar{z}=1,10$ and $\d z=0.1,3$ as illustrated.
 }
	\label{fig:DMfluxratio}
\end{center}
	\vspace{-4mm}
\end{figure} 

We illustrate the relativistic axion flux considering reference parameters (see Sec.~\ref{ssec:sources}), normalized to the local cold axion DM expectation as in Eq.~\eqref{eq:phiscale_DM}, in Fig.~\ref{fig:DMfluxratio}. Here, for our estimates we assume a DM fraction $\Fcal=0.1$, and a Gaussian emission spectrum with a central energy $\bar{\o}=3m$ and with a width $\d\o=m$ (dashed) or $\d\o=m/10$ (thick). The left panel displays the results for a power-law distribution of sources with an exponent $p=1$ (red) or $p=-1$ (purple), and $z_{\rm max}=5,20$. 
For comparison, we also show the result for a Gaussian distribution with $\d z=0.1$ (red) or $\d z=3$ (purple), and $\bar{z}=1,10$ as labeled.
We note that the D$a$B flux is within a few orders of magnitude of the flux expected from the local axion DM density, matching at the order-of-magnitude level to the rough estimates of Eq.~\eqref{eq:fluxfrac}, with a spread in energy that depends on the $\d\o$ as well as the cosmological history encoded in $R_{\rm burst}(z)$.

Fig.~\ref{fig:DMfluxratio} illustrates how the parameters of the cosmological rate function and burst spectrum determine the present-day flux in the D$a$B. For example, in the left panel (power-law rate function), extending the duration of the burst to higher redshifts, from $z_{\rm max}=5$ (faint lines) to $z_{\rm max}=20$ (opaque lines) leads to a more significant low-energy component. The power-law exponent $p$, in contrast, has a more subtle effect on the present-day spectrum. 
For a Gaussian rate function, the peak energy after redshifting is found near $\bar{\o}/(1+\bar{z})$, 
and when the Gaussian width in redshift is smaller more of the energy remains near the peak, as in the red lines ($\d z=0.1$) compared to the purple ones ($\d z=3$). As we have demonstrated, at high redshift the DM density is higher than today, allowing a D$a$B flux larger than the DM background for example in case of Gaussian distribution at high redshift (shown by faint lines in the right panel of Fig.~\ref{fig:DMfluxratio}).
In both panels, the dashed lines, which correspond to a larger width in energy $\d\o$, spread significantly in energy away from the initial peak at $\bar{\o}=3m$ than the solid lines, which exhibit a sharper cutoff. 

Our analysis indicates that D$a$B flux could be sub-leading, and in some scenarios even dominate, compared to that of local cold axion DM.  
Successful detection of D$a$B flux would prove particularly valuable, providing unique information regarding the evolution of the axion field as well as its SM interactions in cosmological and astrophysical contexts dramatically distinct from cold DM. 
We further comment on direct-detection possibilities for D$a$B in Sec.~\ref{ssec:directdetection}.

\section{Axion conversion to photons}
\label{sec:axionphoton}

Complementary to direct detection, the D$a$B can can be probed through indirect signatures due to axion interactions with SM constituents.
Prominent signals appear from axion conversion through axion-photon interaction of Eq.~\eqref{eq:Lphigg}, either in the presence of magnetic fields or through axion decays. We calculate the rates of axion-photon conversion for each scenario. The resulting photon signals can be searched for in a wide range of telescope and satellite experiments outlined in Sec.~\ref{sec:signals}.
Additional detectable D$a$B signatures could arise from other axion couplings as well, as we briefly discuss in Sec.~\ref{sec:additional}.

\subsection{Magnetic fields}

Propagating axions, including those contributing to D$a$B from various sources as discussed in the previous section, can be efficiently converted (oscillate) to photons $a\to\gamma$ by astrophysical magnetic fields~\cite{Mirizzi:2006zy}. We shall subsequently focus on Milky Way Galactic magnetic fields. 
Consider axions traveling a distance $R$ in a homogeneous magnetic field with a transverse component in the plane normal to direction of propagation $B_T$, which is of order $\m{\rm G}$ in the Milky Way~\cite{Beck:2008ty}. We assume Milky Way's magnetic field to be $B_T = 1$~$\m{\rm G}$ throughout. Then, the probability of conversion is~\cite{Raffelt:1987im}
\begin{equation} \label{eq:Pgphi}
    P_{\gamma \to a} = \left(\D_{a\g}R\right)^2\frac{\sin^2(\D_{\rm osc}R/2)}{(\D_{\rm osc}R/2)^2}~,
\end{equation}
where the oscillation wavenumber is $\D_{\rm osc} \equiv \sqrt{(\D_a-\D_{\rm pl})^2 + 4\D_{a\g}^2}$. We have considered the ultra-relativistic approximation, $\o\gg m$, which we expect to be a generally reasonable approximation\footnote{In other energy regimes, we expect additional corrections.}. There are three distinct contributions to the conversion probability:
\begin{align}
    \D_{a\g} &\equiv \frac{g_{a\g} B_T}{2} 
        \simeq 1.5\cdot10^{-4}\left(\frac{g_{a\g}}{10^{-12}\,{\rm GeV}^{-1}}\right)\left(\frac{B_T}{\mu{\rm G}}\right) 
        {\rm kpc}^{-1}\,, \label{eq:Dphig} \\
    \D_a &\equiv - \frac{m^2}{2\omega} 
        \simeq -7.8\cdot 10^{13}\left(\frac{m}{10^{-11}\,{\rm eV}}\right)^2\left(\frac{10^{-10}\,{\rm eV}}{\o}\right)
        {\rm kpc}^{-1}\,, \label{eq:Dphi} \\
    \D_{\rm pl} &\equiv -\frac{\o_{\rm pl}^2}{2\o} 
        \simeq -1.1\cdot10^{13}\left(\frac{n_e}{10^{-2}\,{\rm cm}^{-3}}\right)\left(\frac{10^{-10}\,{\rm eV}}{\o}\right)
        {\rm kpc}^{-1}\,, \label{eq:Dpl}
\end{align}
where $\o_{\rm pl} = \sqrt{4\pi\a n_e/m_e} \simeq 4\cdot 10^{-11} {\rm eV}\sqrt{n_e/{\rm cm}^{-3}}$ is the plasma frequency corresponding to effective mass that photons acquire induced by the SM background, with $m_e$ being the electron mass and $n_e$ density of free electrons in medium. 

\begin{figure*}
\centering
\begin{minipage}[c]{0.48\textwidth}
\hspace{-0cm} \large{(a)} \\
\vspace{-0.5cm}
\includegraphics[width=\linewidth]{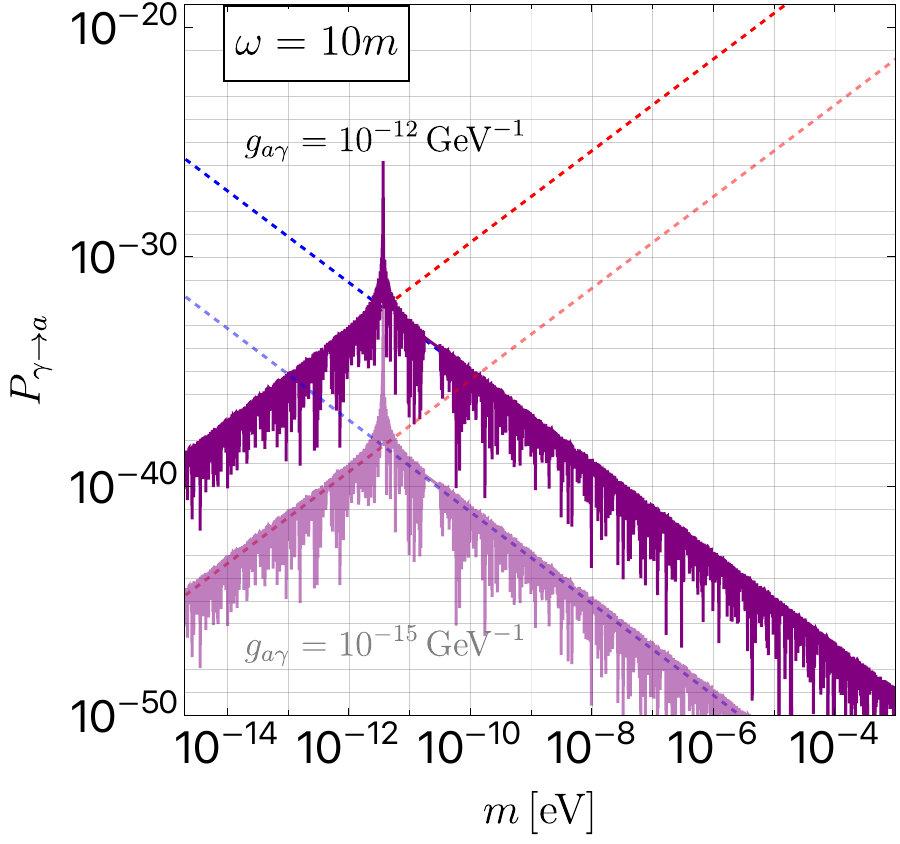}
\end{minipage}
\hfill
\begin{minipage}[c]{0.48\textwidth}
\hspace{-0cm} \large{(b)} \\
\vspace{-0.5cm}
\includegraphics[width=\linewidth]{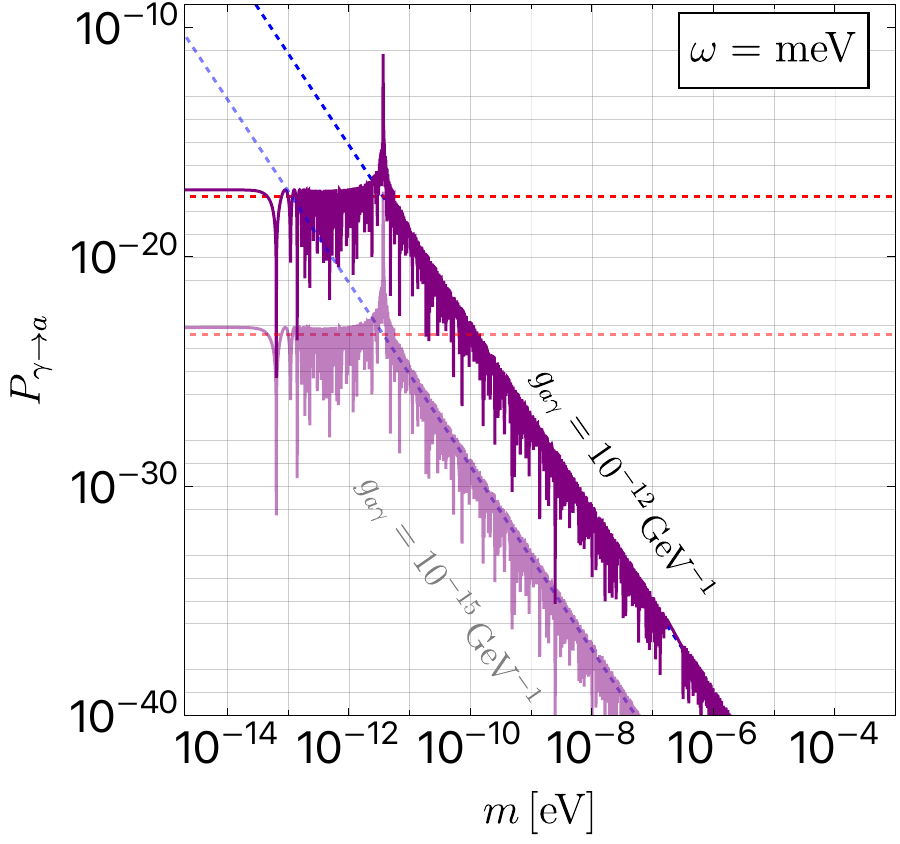}
\end{minipage}
\\
\vspace{0.2cm}
\begin{minipage}[c]{0.48\textwidth}
\hspace{-0cm} \large{(c)} \\
\vspace{-0.5cm}
\includegraphics[width=\linewidth]{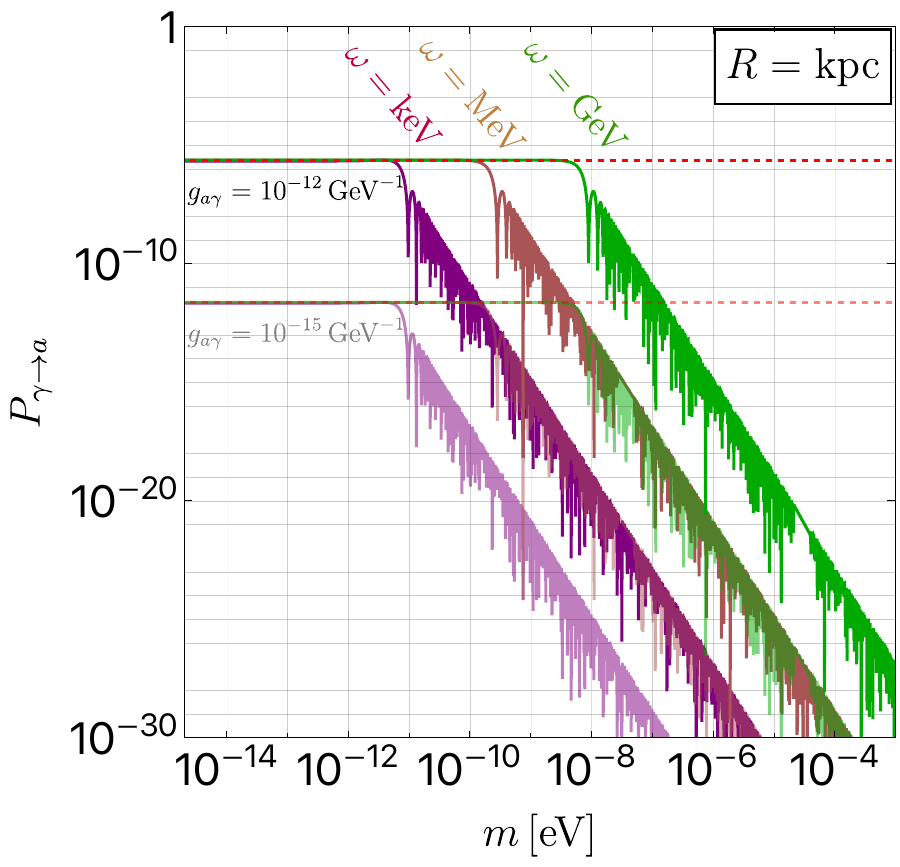}
\end{minipage}
\hfill
\begin{minipage}[t]{.48\textwidth}
\vspace{-3cm}
\caption{\label{fig:Pphigamma}
Conversion probability for axion to photon $a\to\gamma$ as a function of axion mass $m$, in the (transverse) Galactic magnetic field $B_T=\mu{\rm G}$ considering an electron density of $n_e=10^{-2}\,{\rm cm}^{-3}$. The probability is computed for fixed $\omega$ equal to $10m$ (panel a), $\rm meV$ (panel b), or $\gg{\rm eV}$ (panel c, energies keV, MeV, GeV as labeled). The purple, brown, and green lines  show the full solutions of Eq.~\eqref{eq:Pgphi}, whereas the dashed red and blue fit lines illustrate the limiting cases at small and large $m$ (see text for details). The axion-photon coupling is fixed in the thick (faint) lines to a value of $g_{a\g} = 10^{-12}\,{\rm GeV}^{-1}$ ($10^{-15}\,{\rm GeV}^{-1}$).
}
\end{minipage}
\end{figure*}

For lower energy emitted axions, when the energy scales could be semi-relativistic with $\o$ are not very far from $m$, we find $\D_{a\g}$ is negligible at all relevant $m$. The other two contributing terms are equal, $\D_a \simeq \D_{\rm pl}$, when $m = \omega_{\rm pl}$. As indicated by Eqs.~(\ref{eq:Dphi}-\ref{eq:Dpl}), both $|\D_a|,|\D_{\rm pl}|\gg {\rm kpc}^{-1}$ in this case. Hence, averaging the resulting fast oscillations in Eq.~\eqref{eq:Pgphi} as $\sin^2\to 1/2$, we can write
\begin{align} \label{eq:Pgphi2}
    P_{\gamma \to a} &\simeq 
        (\D_{a\g}R)^2 \dfrac{\sin^2(|\D_a-\D_{\rm pl}|R/2)}{(\D_a-\D_{\rm pl})^2R^2/4} \nn \\
        &\simeq 
    \begin{cases}
        \displaystyle 2\left(\dfrac{\D_{a\g}}{\D_{\rm pl}}\right)^2 \\
        \displaystyle 2\left(\dfrac{\D_{a\g}}{\D_{a}}\right)^2   
    \end{cases}\\ 
        &= 
        \begin{cases}
         2\cdot 10^{-32}\left(\dfrac{g_{a\g}}{10^{-12}\,{\rm GeV}^{-1}}\right)^2
        \left(\dfrac{\o}{10^{-10}\,{\rm eV}}\right)^2
        \left(\dfrac{10^{-2}\,{\rm cm}^{-3}}{n_e}\right)^2
        \left(\dfrac{B_T}{\mu{\rm G}}\right)^2
        \,,& m \lesssim \omega_{\rm pl}\,, \\
         4\cdot10^{-34}\left(\dfrac{g_{a\g}}{10^{-12}\,{\rm GeV}^{-1}}\right)^2
        \left(\dfrac{\o}{10^{-10}\,{\rm eV}}\right)^2
        \left(\dfrac{10^{-11}\,{\rm eV}}{m}\right)^4
        \left(\dfrac{B_T}{\mu{\rm G}}\right)^2\,,& m \gtrsim \omega_{\rm pl}~. \notag
    \end{cases}
\end{align}
Note that the result is essentially independent of $R$, since the argument of the sine function in Eq.~\eqref{eq:Pgphi} is significantly larger than unity for all astrophysical, and certainly cosmological, distances. 
In Fig.~\ref{fig:Pphigamma}, we illustrate the two limits of Eq.~\eqref{eq:Pgphi2} (red and blue curves, respectively), and compare to the full expression of Eq.~\eqref{eq:Pgphi} (purple). As indicated, the conversion probability is proportional to $P_{\gamma \rightarrow a} \propto g_{a\g}^{2}\o^2$.

In the other limiting regime, 
when $m\ll 10^{-11}$ eV and $\o \gg $MeV (as in e.g. environment of supernovae), the first term dominates and $P_{\g\to a}$ becomes independent of $m$ and $\o$~\cite{Mirizzi:2006zy,Calore:2020tjw}. In that case
\begin{equation}
    P_{\g\to a} \simeq (\D_{a\g}R)^2
        \simeq 2\cdot10^{-6}\left(\frac{g_{a\g}}{10^{-12}\,{\rm GeV}^{-1}}\right)^2
            \left(\frac{B_T}{\m{\rm G}}\right)^2
            \left(\frac{R}{\rm kpc}\right)^2\,,
\end{equation}
which approaches unity $P_{\g\to a}\to 1$ for sizable distances $R$. In panel (c) of Fig.~\ref{fig:Pphigamma} we illustrate the conversion probability for $R=$~kpc and for different axion energies $\o=$ keV, MeV, and GeV (purple, brown, and green lines respectively). 

Using the above, in Sec.~\ref{sec:signals} below we estimate the sensitivity of experiments to detect photon signals generated by axion-photon conversion. For this, the flux of photons can be readily computed from D$a$B flux as
\begin{align} \label{eq:dphido_Bfield}
\frac{d\phi_\g}{d\o}\Big|_{B{\rm -field}} = 
        P_{\g\to a}\frac{d\phi}{d\o}
\end{align}    
where the axion flux $d\phi/d\o$ is given by Eq.~\eqref{eq:dphido_Gaus}. In what follows, we consider characteristic $R=1\,{\rm kpc}$, as the typical coherence length for Galactic magnetic fields is $\Ocal({\rm few})\,{\rm kpc}$~\cite{Beck:2008ty}. 

\begin{figure} 
\begin{center}
  \includegraphics[width=1\textwidth]{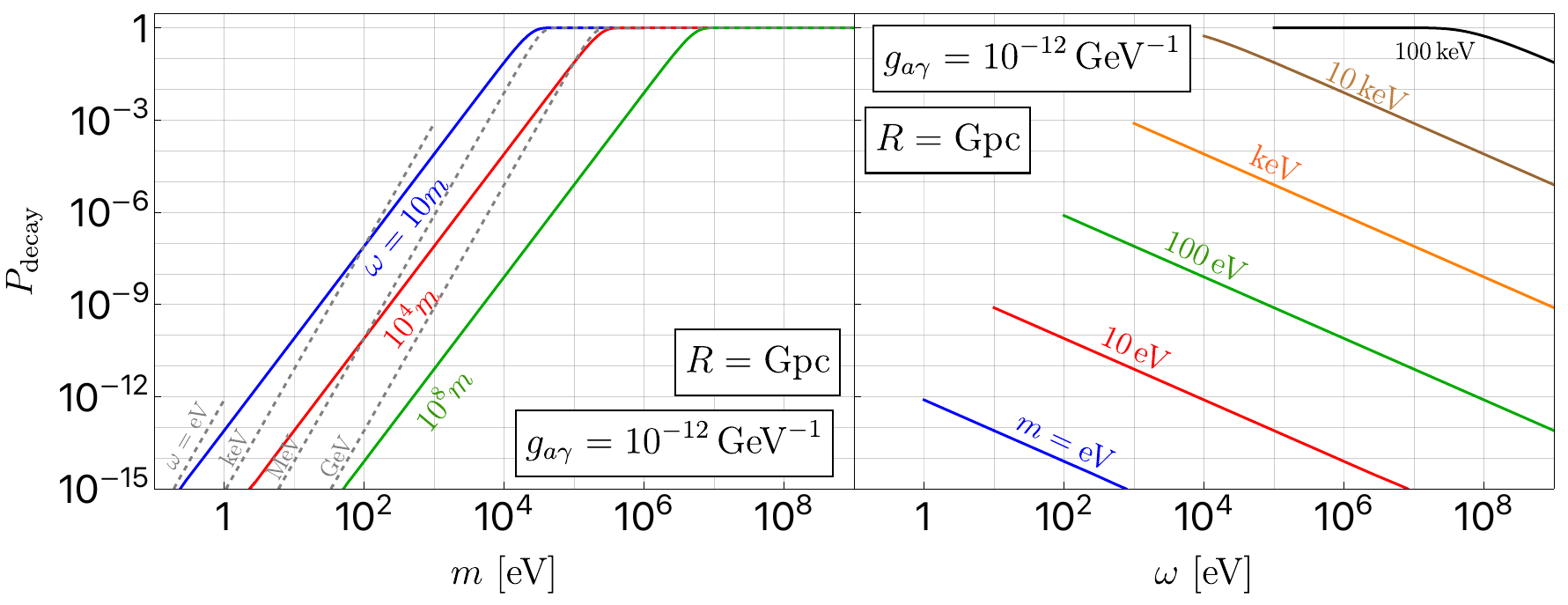}  
\caption{Axion-photon decay probability factor, $P_{\rm decay}$, as a function of axion mass $m$ at fixed energy $\o$ (left panel), and as a function of axion energy $\o$ at fixed mass $m$ (right panel). Representative values of axion coupling $g_{a\g}=10^{-12}\,{\rm GeV}^{-1}$ and cosmological propagation distance $R={\rm Gpc}$ are considered. }
	\label{fig:Pdecay}
\end{center}
	\vspace{-4mm}
\end{figure} 

\subsection{Axion decay}

Axions can decay to photons, $a\to\g\g$, through the axion-photon coupling in Eq.~\eqref{eq:Lphigg}. The decay rate of an axion in vacuum is given by
\begin{equation}
    \G_{\g\g} = \frac{g_{a\g}^2m^3}{64\pi}~.
\end{equation}
Hence, axions of energy $\o$ traveling a distance $R$ will decay with probability $P_{\rm decay}(R,\o) = 1-\exp(-R/\ell(\o))$, where the mean free path is given by
\begin{equation} \label{eq:decaylength}
    \ell(\o) \simeq \frac{\g v_{\rm burst}}{\G_{\g\g}}
        \simeq \left(\frac{\o}{m}\right)
            \frac{64\pi}{g_{a\g}^2m^3}
        \simeq {\rm Mpc}\left(\frac{\o}{{\rm MeV}}\right)\left(\frac{100\,{\rm keV}}{m}\right)^4
            \left(\frac{10^{-12}\,{\rm GeV}^{-1}}{g_{a\g}}\right)^{2}\,,
\end{equation}
with $\g$ is the relativistic Lorentz factor. 
As Eq.~\eqref{eq:decaylength} illustrates, axions in the D$a$B can travel cosmological distances before decaying, and therefore the photon flux from decay must be computed self-consistently over cosmological redshifts $z$ as
\begin{equation} \label{eq:dphigdo_decay}
    \ddx{\phi_\g}{\omega}\Big|_{\rm decay} = \int_0^\infty dz 
        (1+z)\ddx{N_a(\omega(1+z))}{\omega} R_{\rm burst}(z)
                \left|\frac{dt}{dz}\right| 2P_{\rm decay}(R(z),\o(1+z)) ~,
\end{equation}
which accounts for the dependence of the source distance $R$ and emission energy $\o$ on redshift $z$.
For a sharply-peaked distribution around some redshift $\bar{z}$, we can treat $P_{\rm decay}$ as a constant in the integration, and the decay flux takes the form
\begin{align} \label{eq:dphigdo_decay_appx}
\frac{d\phi_\g}{d\o}\Big|_{\rm decay} \simeq 2P_{\rm decay}\left(R(\bar{z}),\o(1+\bar{z})\right)\frac{d\phi}{d\o}~,
\end{align}
which is analogous to Eq.~\eqref{eq:dphido_Bfield} for magnetic field conversion. 
 Note that there is an additional factor of 2 compared to magnetic field conversion case, since each axion decay creates two photons.

For example for Galactic contributions to D$a$B, such as from astrophysical supernovae, the distances to the sources $R$ could be related to their astrophysical population distributions, e.g. localized around Galactic Center.
More generally, particularly for dark sector sources, the source distances $R(z)$ can be cosmological in scales, e.g. for $z\gtrsim 1$ we have $R(z)\gtrsim {\rm Gpc}$, but depend in detail on the population distribution of axion emission sources. We define an effective redshift for the purpose of estimating the typical source distance in generality, $z_{\rm eff}$, as
\begin{equation} \label{eq:zeff}
    z_{\rm eff} \equiv \dfrac{\int dt \dfrac{z\,R_{\rm burst}(z)}{(1+z)^3}}
    {\int dt \dfrac{R_{\rm burst}(z)}{(1+z)^3}}\,.
\end{equation}
For a peaked distribution $R_{\rm burst}(z)$, such as the Gaussian introduced in Eq.~\eqref{eq:fzmodels}, Eq.~\eqref{eq:zeff} gives $z_{\rm eff}\simeq\bar{z}$ to a good approximation if $\bar{z}\gtrsim 5$. For the power-law distribution in Eq.~\eqref{eq:fzmodels}, which can be broader in redshift, $z_{\rm eff}$ is somewhat smaller, and we can approximate $z_{\rm eff} \simeq 2/(7-2p)$. To estimate the decay probability in what follows, we use $R(z)\simeq 1/H(z_{\rm eff})$ where $H(z)$ is the Hubble parameter.

As noted by~Ref.~\cite{Calore:2020tjw}, for large axion masses and energies $\o,m\gtrsim{\rm MeV}$ (appropriate for e.g. heavy axions produced in supernovae), nearly all axions are expected to decay on cosmological distances, as the exponential in Eq.~\eqref{eq:dphigdo_decay_appx} approaches zero, or equivalently when $R/\ell \gg 1$. On the other hand, when $R/\ell \ll 1$, we can approximate the flux as $d\phi_\g/d\o \simeq (2 R/\ell)d\phi/d\o$, which is proportional to $\o^{-1}m^{4}$ multiplied by the axion flux. We illustrate the behavior of $P_{\rm decay}$ 
in different limits in Fig.~\ref{fig:Pdecay}.

\section{Observable signatures with photons}
\label{sec:signals}

The generation of photon flux from D$a$B, either through magnetic $B$-field conversion or axion decays, allows to search for axions over a broad parameter space with a range of detectors and telescopes. In principle, the contributions from these processes are additive, stemming from the same axion-SM interaction $g_{a\g}$. However, as we demonstrated, the two processes are expected to contribute a significant flux of photons in distinct kinematic regimes. Approximately, axion magnetic field conversion tends to dominate at high energies when $\o\gg m$
whereas axion decays are more prominent when their energy $\o$ is within a few orders of magnitude of the mass $m$. 

\begin{figure} 
\begin{center}
  \includegraphics[width=1\textwidth]{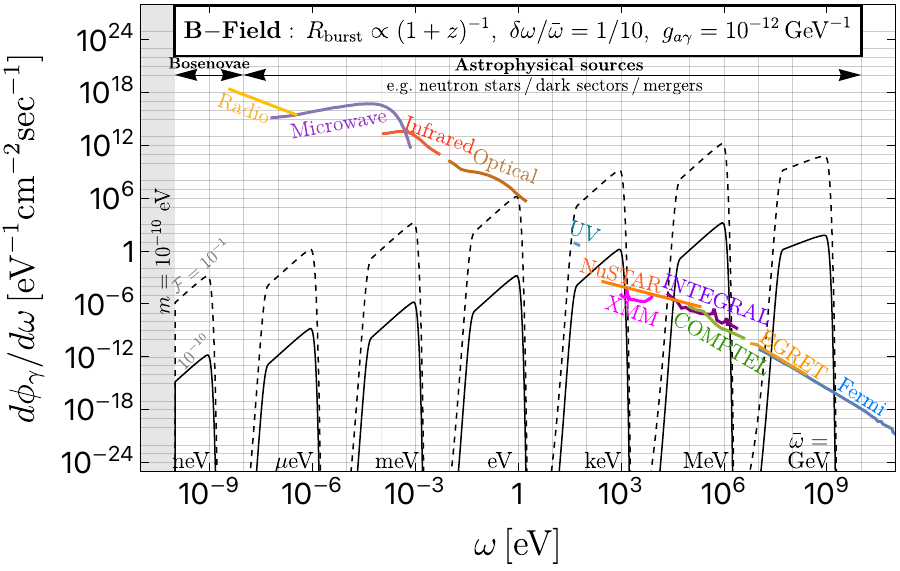}  
	\caption{
 Photon flux from Galactic magnetic $\mu$G $B$-field conversion of the D$a$B (black lines) assuming $\Fcal=0.1$ (dashed) or $10^{-10}$ (thick), for axions with mass $m=10^{-10}\,{\rm eV}$ and coupling $g_{a \gamma} = 10^{-12}$~GeV$^{-1}$ emitted from transient sources with cosmological burst rate $R_{\rm burst} \propto (1+z)^{-1}$, ratio of central energy spread to average magnitude $\delta w/\bar{\o} = 1/10$. Each pair of black lines corresponds to burst sources with central energy $\bar{\o}$ as labeled along the x-axis. Contributing background fluxes for a wide range of photon frequencies (from radio, microwave, infrared, optical, and UV) are reproduced from~Ref.~\cite{Cooray:2016jrk} (see also references therein). At higher energies $\o\gtrsim 100\,{\rm eV}$, we display the photon flux measured by specific experiments, which lead to constraints on D$a$B, including 
    NuSTAR~\cite{NuSTAR:2013yza,Roach:2022lgo}, 
    XMM-Newton~\cite{Foster:2021ngm}, 
    INTEGRAL~\cite{Bouchet:2008rp}, 
    COMPTEL~\cite{2008ICRC....3.1085S,Strong:2019yfj}, 
    EGRET~\cite{Strong:2004de}, and 
    Fermi-LAT~\cite{Fermi-LAT:2014ryh,
    Principe:2022ttq}.
}
	\label{fig:summary}
\end{center}
	\vspace{-4mm}
\end{figure}

In Fig.~\ref{fig:summary} and Fig.~\ref{fig:summary_decay}, we illustrate the photon flux obtained from Galactic magnetic $B$-field conversion as well as axion decays, respectively. For comparison, we include the measured background flux from telescope experiments across a wide range of frequencies, from gamma-ray to radio, spanning nearly 20 orders of magnitude in photon energy. 
We observe that over a very broad parameter space, the photon flux obtained from D$a$B can greatly exceed other contributing backgrounds, allowing to impose constraints and promising future detection prospects.

Highlighting our results, we estimate in detail the sensitivity of several distinct experiments below. For concreteness, we consider a power-law distribution of cosmological transient sources $R_{\rm burst}\propto (1+z)^p$ taking $p=-1$ and assuming a maximum redshift of $z_{\rm max}=20$.
Computations of results for the case of power-law distribution of cosmological transient sources but with $p=1$, as well as a Gaussian distribution, are depicted in App.~\ref{app:RateFunc}. We find the results for these cases to be overall similar to those presented below for power-law distribution with $p = -1$, suggesting that dependence on the detailed cosmological assumptions about transient source population is not very strong. Further, we find that the spread of the axion energy distribution, $\d\o/\bar{\o}$, determines the position of the low-energy cutoff of each black curve. 

\subsection{Limits from X-ray and gamma-ray searches}
\label{ssec:FermiComptel}
\begin{figure}
\begin{center}
  \includegraphics[width=1\textwidth]{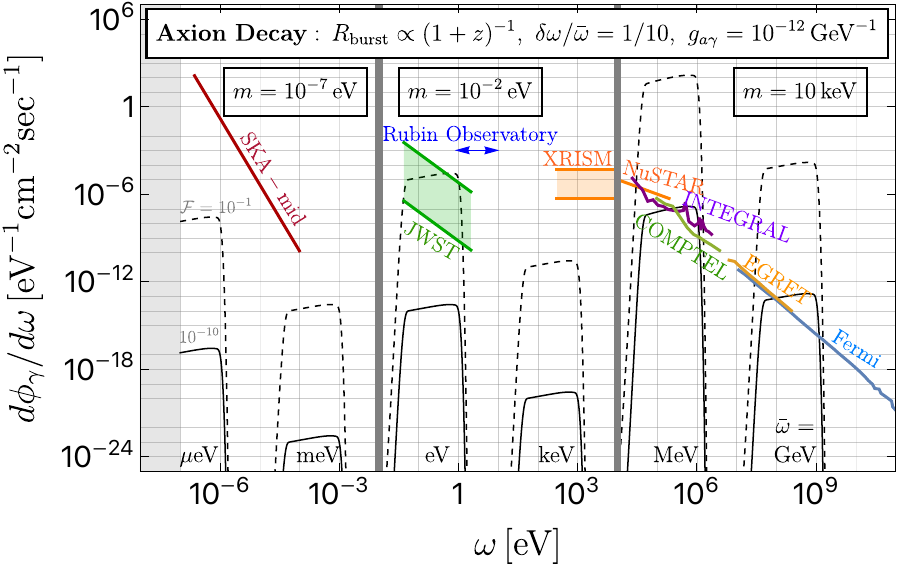}  
	\caption{Photon flux from decay of axions from D$a$B (black lines) assuming $\Fcal=0.1$ (dashed) or $10^{-10}$ (thick), considering axion coupling $g_{a\gamma} = 10^{-12}$~GeV$^{-1}$ and mass $m=10^{-7}\,{\rm eV}$ (left), $10^{-2}\,{\rm eV}$ (center), and $10\,{\rm keV}$ (right). 
 D$a$B is considered to arise from transient sources with cosmological burst rate of $R_{\rm burst} \propto (1+z)^{-1}$, assuming typical source distance of $R(z_{\rm eff})\simeq 1/H(z_{\rm eff})$ with $z_{\rm eff}\simeq 2/9$. Each pair of black lines corresponds to transient sources with central energy $\bar{\o}$ as labeled along the x-axis. (right) Photon fluxes measured by specific experiments and leading to constraints on D$a$B are as described in the caption of Fig.~\ref{fig:summary}. At lower energies (below MeV, center, and below eV, left), we illustrate sensitivity of future searches by Vera C. Rubin Observatory (formerly LSST)~\cite{Tyson:2002mzo,Mao:2022fyx}, 
JWST~\cite{Bessho:2022yyu,JWST:2023abc,Janish:2023kvi,Roy:2023omw}, 
XRISM~\cite{XRISMScienceTeam:2020rvx} and SKA~\cite{Weltman:2018zrl,Braun:2019gdo,Ghosh:2020hgd} as described in Sec.~\ref{ssec:SKA} (see also Fig.~\ref{fig:summary_future}).}
	\label{fig:summary_decay}
\end{center}
	\vspace{-4mm}
\end{figure} 

Variety of experiments can sensitively probe flux of energetic photons with $\o\gg {\rm eV}$ from D$a$B. We illustrate this in Fig.~\ref{fig:summary} and Fig.~\ref{fig:summary_decay}, for selected input parameters, using the measured photon background flux of several existing experiments including      
    NuSTAR~\cite{NuSTAR:2013yza,Roach:2022lgo}, 
    XMM-Newton~\cite{Foster:2021ngm}, 
    INTEGRAL~\cite{Bouchet:2008rp}, 
    COMPTEL~\cite{2008ICRC....3.1085S,Strong:2019yfj}, 
    EGRET~\cite{Strong:2004de}, and 
    Fermi-LAT~\cite{Fermi-LAT:2014ryh,Principe:2022ttq}.
As a characteristic example of Fermi-LAT, we consider 8-yr dataset observations of isotropic diffuse gamma-ray background\footnote{ULTRACLEANVETO event class, with reduced cosmic-ray contamination. See  iso\underline{ }P8R3\underline{ }ULTRACLEAN\underline{ }V3\underline{ }v1.txt at \href{https://fermi.gsfc.nasa.gov/ssc/data/access/lat/BackgroundModels.html}{https://fermi.gsfc.nasa.gov/ssc/data/access/lat/BackgroundModels.html}.}, which above can be modeled using a simple fit function~\cite{Calore:2020tjw}
\begin{equation}
    \frac{d\phi_\g(\o)}{d\o}\Bigg|_{\rm Fermi} \simeq 5.5\cdot10^{-10}\left(\frac{\rm GeV}{\o}\right)^{2.2}\,{\rm MeV}^{-1}{\rm cm}^{-2}{\rm sec}^{-1}
\end{equation}
in the $\omega$ energy range $30\,{\rm MeV}-300\,{\rm GeV}$. Analogously, COMPTEL observations~\cite{2008ICRC....3.1085S} of diffuse gamma-ray emission can be fit as~\cite{Calore:2020tjw}
\begin{equation}
    \frac{d\phi_\g(\o)}{d\o}\Bigg|_{\rm COMPTEL}\simeq 5\cdot10^{-3}\left(\frac{{\rm MeV}}{\o}\right)^{2.4}{\rm MeV}^{-1}{\rm cm}^{-2}{\rm sec}^{-1}
\end{equation}
in the $\omega$ energy range $0.8-30\,{\rm MeV}$. We summarize observations employed in our analysis in Tab.~\ref{tab:labs}.
Overall, we analyze combined observations in the $\omega$ energy range of approximately ${\rm MeV}-{\rm 300~GeV}$ showcasing their implications for D$a$B, and leave a more comprehensive study of signals outside of this range to future work. Other measured fluxes, including those of EGRET, INTEGRAL, XMM, NuSTAR, and low-energy telescopes, are summarized in Tab.~\ref{tab:labs}.

\begin{table}[tb]
\begin{center}
\begin{tabular}{l | rccc}
\hline\hline
 Experiment(s) & Energy Range & Flux  $(d\phi_\g/d\o)$ Fit [eV$^{-1}$cm$^{-2}$sec$^{-1}$] \\
 \hline
  Radio*~\cite{Cooray:2016jrk} & $4-340\,{\rm neV}$ & $6.2\cdot10^{14}\left(\dfrac{\mu{\rm eV}}{\o}\right)^{1.5}$ \\
  Microwave*~\cite{Cooray:2016jrk} & $67\,{\rm neV} - 730\,\mu{\rm eV}$ & $3.2\cdot10^{17}\left(\dfrac{\o}{\rm meV}\right)^{0.5}\exp\left(-\dfrac{15\o}{\rm meV}\right)$ \\
  Infrared*~\cite{Cooray:2016jrk} & $0.1-5\,{\rm meV}$ & $6.3\cdot10^{13}\left(\dfrac{\o}{\rm meV}\right)^{0.5}\exp\left(-\dfrac{1.6\o}{\rm meV}\right)$ \\
  Optical*~\cite{Cooray:2016jrk} & $10\,{\rm meV} - 1.7\,{\rm eV}$ & $4.4\cdot10^6\left(\dfrac{\rm eV}{\o}\right)^{1.8}$ \\
  UV*~\cite{Cooray:2016jrk} & $46-58\,{\rm eV}$ & $1.7\cdot10^{4}\left(\dfrac{\rm eV}{\o}\right)^{2}$ \\
  NuSTAR~\cite{NuSTAR:2013yza,Roach:2022lgo} & $0.3-200\,{\rm keV}$ & $10^{-4} \left(\dfrac{\rm keV}{\o}\right)$ \\
  XMM-Newton~\cite{Foster:2021ngm} & $1-8\,{\rm keV}$ & $8.9\cdot10^{-6}\left(\dfrac{\rm keV}{\o}\right)^{0.65}$ \\
  INTEGRAL~\cite{Bouchet:2008rp} & $24\,{\rm keV} - 2\,{\rm MeV}$ & $10^{-8} \left(\dfrac{\rm MeV}{\o}\right)^{1.6}$ \\
  COMPTEL~\cite{2008ICRC....3.1085S,Strong:2019yfj} & $0.8-30\,{\rm MeV}$ & $5\cdot10^{-9}\left(\dfrac{{\rm MeV}}{\o}\right)^{2.4}$ \\
  EGRET~\cite{Strong:2004de} & $6-240\,{\rm MeV}$ & $4\cdot 10^{-16}\left(\dfrac{\rm GeV}{\o}\right)^{2.3}$ \\
  Fermi-LAT~\cite{Fermi-LAT:2014ryh,Principe:2022ttq} & $30\,{\rm MeV}-300\,{\rm GeV}$ & $5.5\cdot10^{-16}\left(\dfrac{\rm GeV}{\o}\right)^{2.2}$ \\
  \hline \hline 
\end{tabular}
\end{center}
  \footnotesize{~~~* Denotes combined observations from multiple experiments~\cite{Cooray:2016jrk}.}
\caption{Summary of experiments whose observations are used to analyze photon signals from D$a$B derived in this work, along with their sensitivity energy range. The fits to the total measured photon flux in the Milky Way by experiments are also displayed. See Fig.~\ref{fig:summary} for further details.  
} 
\label{tab:labs}
\end{table}

\begin{figure} 
\begin{center}
  \includegraphics[width=0.9\textwidth]{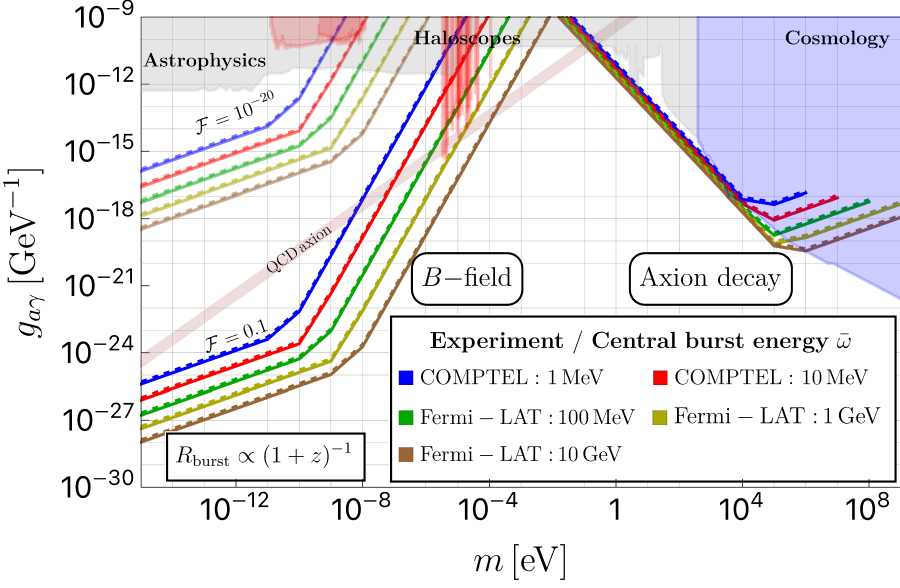}  
\caption{Sensitivity to D$a$B photons of COMPTEL (blue, red) and Fermi-LAT (green, yellow, and brown), as a function of axion mass $m$ for average transient burst energy $\bar{\o}$. The emitted photons are produced from either Galactic magnetic $\mu$G $B$-field conversion (left side) or axion decays (right). D$a$B is considered to arise from transient sources with cosmological burst rate of $R_{\rm burst} \propto (1+z)^{-1}$. For $B$-field conversion we take $R={\rm kpc}$, and for axion decay we assume a typical source distance of $R(z_{\rm eff})\simeq 1/H(z_{\rm eff})$ with $z_{\rm eff}\simeq 2/9$. The upper (lower) lines correspond to $\Fcal=0.1$ ($\Fcal=10^{-20}$).  
The thick (dashed) lines correspond to transient burst with energy spread $\d\o = \bar{\o}/10$ ($\d\o = \bar{\o}$). For enumeration of astrophysical limits (gray) below $m\lesssim 0.1~{\rm eV}$, see Fig.~\ref{fig:sens}. At $m\gtrsim 0.1\,{\rm eV}$ we also include astrophysical limits (gray) from the Hubble Space Telescope (HST)~\cite{Carenza:2023qxh}, James Webb Space Telescope (JWST)~\cite{Janish:2023kvi}, the Visible Multi-Object Spectrograph (VIMOS)~\cite{Grin:2006aw}, MUSE-Faint survey~\cite{Todarello:2023hdk}, and the Breakthrough Listen Galactic Center Survey~\cite{Foster:2022fxn},  as well as direct DM detection searches using haloscopes (red) including ADMX~\cite{ADMX:2018gho,ADMX:2018ogs,ADMX:2019uok,Crisosto:2019fcj,ADMX:2021mio,ADMX:2021nhd}, 
CAPP~\cite{Lee:2020cfj,Jeong:2020cwz,CAPP:2020utb,Lee:2022mnc,Kim:2022hmg,Yi:2022fmn,Yang:2023yry},
CAST-CAPP~\cite{Adair:2022rtw},
CAST-RADES~\cite{CAST:2020rlf},
GrAHal~\cite{Grenet:2021vbb}, 
RBF~\cite{DePanfilis}, HAYSTAC~\cite{HAYSTAC:2018rwy,HAYSTAC:2020kwv,HAYSTAC:2023cam}, QUAX~\cite{Alesini:2019ajt,Alesini:2020vny,Alesini:2022lnp},  ORGAN~\cite{McAllister:2017lkb,Quiskamp:2022pks}, and
TASEH~\cite{TASEH:2022vvu}. For $m\gtrsim 100\,{\rm eV}$ we illustrate  leading constraints from cosmology (blue) from Ref.~\cite{Cadamuro:2011fd}.}
	\label{fig:sens2}
\end{center}
	\vspace{-4mm}
\end{figure}

In order to obtain constraints on axion interactions,
we vary the coupling $g_{a\g}$ to derive the sensitivity regions for Fermi-LAT and COMPTEL, highlighting our analysis, where the peak photon fluxes from the D$a$B exceeds the gamma-ray fluxes displayed in Tab.~\ref{tab:labs}. We stress that our analysis can be readily analogously carried out for other experimental observations (see e.g. Tab.~\ref{tab:labs}).
The resulting limits and constraints are illustrated in Fig.~\ref{fig:sens2}. The lines correspond to searches over the range of energies given in the legend, for COMPTEL (blue and red lines) and Fermi-LAT (green, yellow, and brown lines). 
As discussed previously, Galactic $B$-field conversion (see Eq.~\eqref{eq:dphido_Bfield}) is most relevant for small axion masses $m\ll {\rm eV}$, whereas decays (see Eq.~\eqref{eq:dphigdo_decay}) dominate at larger axion masses. The change in the slope of the lines correspond to shifts in the behavior of the photon conversion and decay probabilities, seen in Fig.~\ref{fig:Pphigamma} and Fig.~\ref{fig:Pdecay}, respectively. For comparison, we also show the band corresponding to the QCD axion in pink~\cite{Kim:1979if,Shifman:1979if,Dine:1981rt,Zhitnitsky:1980tq}. 

The total flux is proportional to $\propto g_{a\g}^2\Fcal$, allowing for simple rescaling of our benchmarks for different DM fraction $\Fcal$. We illustrate in Fig.~\ref{fig:sens2} scenarios of $\Fcal=0.1$ and $\Fcal=10^{-20}$.
We observe that even for a very small converted DM fraction $\Fcal\lesssim 10^{-20}$, we find the estimated sensitivity for axions searches converting to photons due to Galactic magnetic fields far exceeds existing constraints from astrophysics or conventional direct DM detection searches using haloscopes (see Fig.~\ref{fig:sens2} caption for details). At large masses, we find signals from axion decays can be competitive with strong existing cosmological limits~\cite{Cadamuro:2011fd}.

We note that our estimates are conservative, since the photon flux associated with D$a$B could be detectable even if it is subdominant to background. Our analysis thus motivates dedicated searches, which would include detailed information about the distribution of D$a$B photons as a function of energy and could potentially significantly further improve these results.

\subsection{Sensitivity of future searches}
\label{ssec:SKA}

Rich opportunities exist to probe photons from D$a$B over a broad range of energies. 
Complementing observations at high energies $\omega \gg$~eV discussed in previous section, at low energies $\o\lesssim {\rm eV}$ there also exist significant backgrounds from radio, microwave, infrared, and optical frequency photons, as summarized in Fig.~\ref{fig:summary}. However, many recently-launched as well as proposed future experiments are expected to probe significantly smaller fluxes over a wide range of energies. These include 
the Square Kilometer Array (SKA) with sensitivity in the radio range $\o<10^{-4}\,{\rm eV}$~\cite{Weltman:2018zrl,Braun:2019gdo,Ghosh:2020hgd}, 
Vera C. Rubin Observatory\footnote{Formerly, Large Synoptic Survey Telescope, LSST.} operating at optical energy range $1-10\,{\rm eV}$~\cite{Tyson:2002mzo,Mao:2022fyx}, 
James Webb Space Telescope (JWST) sensitivity in the infrared of $0.04-2\,{\rm eV}$~\cite{Bessho:2022yyu,JWST:2023abc,Janish:2023kvi,Roy:2023omw}, 
recently launched XRISM with sensitivity for X-rays of $0.3-12\,{\rm keV}$~\cite{XRISMScienceTeam:2020rvx}, 
as well as the e-ASTROGAM and AMEGO proposals with sensitivity for gamma rays in the range of $0.1\,{\rm MeV}-300\,{\rm MeV}$, estimated as having similar sensitivity~\cite{e-ASTROGAM:2017pxr,DeAngelis:2021abc,AMEGO:2021abc,Engel:2022bgx}. 

\begin{figure} 
\begin{center}
  \includegraphics[width=1\textwidth]{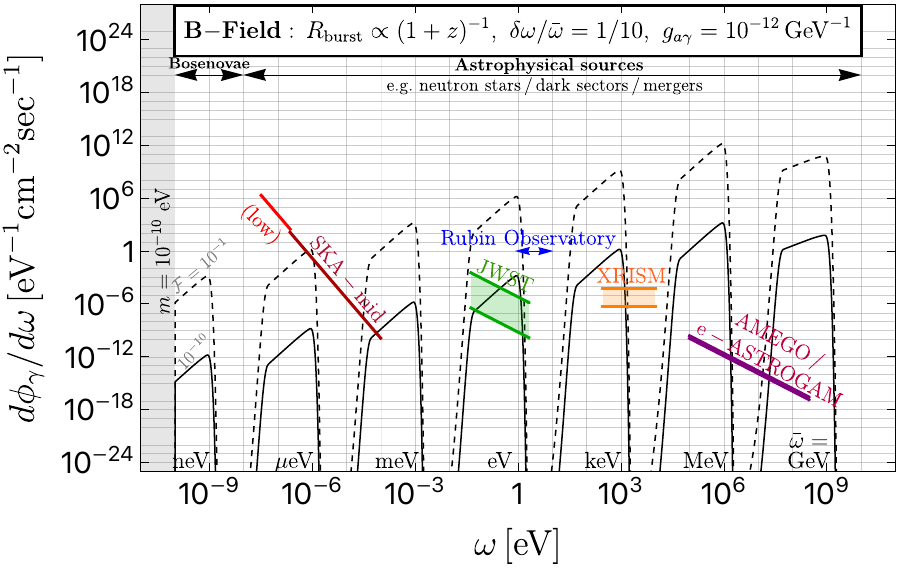}  
	\caption{Same as Fig.~\ref{fig:summary}, but illustrating the expected sensitivity of recently-launched and future experiments (see App.~\ref{app:photonest} and references for flux estimation) including SKA~\cite{Weltman:2018zrl,Braun:2019gdo,Ghosh:2020hgd},
 Vera C. Rubin Observatory (formerly LSST)~\cite{Tyson:2002mzo,Mao:2022fyx}, 
JWST~\cite{Bessho:2022yyu,JWST:2023abc,Janish:2023kvi,Roy:2023omw}, 
XRISM~\cite{XRISMScienceTeam:2020rvx}, and 
gamma-ray proposals e-ASTROGAM and AMEGO~\cite{e-ASTROGAM:2017pxr,DeAngelis:2021abc,AMEGO:2021abc,Engel:2022bgx}. Note that the blue arrows labeled ``Rubin Observatory'' denote only a range of sensitivity in $\o$, not a photon flux sensitivity estimate.}
	\label{fig:summary_future}
\end{center}
	\vspace{-4mm}
\end{figure}

We estimate, see App.~\ref{app:photonest} for details, the expected photon background flux observations from experiments and illustrate comparison with the D$a$B flux in Fig.~\ref{fig:summary_future}.
We note that precise sensitivity estimates could depend on a multitude of relevant factors, such as the source, telescope direction, noise estimates as well as duration. 

To showcase different detection approaches, we highlight the sensitivity estimate of SKA to the flux of photons from D$a$B.\footnote{SKA analysis has also been performed to search for decay of cosmological DM to axions, $\chi\to aa$, which subsequently convert to photons~\cite{Kar:2022ngx}.} SKA is expected to have unprecedented sensitivity reach at low radio energies $\omega < 10^{-4}$~eV and the methodology is qualitatively distinct from that of searches with high-energy gamma-ray telescopes, such as Fermi-LAT as discussed in previous Sec.~\ref{ssec:FermiComptel}. Expected performance of SKA can be found in Ref.~\cite{Braun:2019gdo}.
We consider two planned realizations of the experiment: \emph{SKA-low}, with a collecting surface area of $A\simeq 419,000\,{\rm m}^2$ and operating in the $50-350~{\rm MHz}$ frequency range, and \emph{SKA-mid}, with a collecting surface area of $A\simeq 10^6\,{\rm m}^2$ (i.e. 5659 antennas with 15~m diameter) and detection range of $350~{\rm MHz}-15.4~{\rm GHz}$. 
For both configurations we assume a detection efficiency of $\eta\simeq 0.8$.

\begin{figure} 
\begin{center}
  \includegraphics[width=0.50\textwidth]{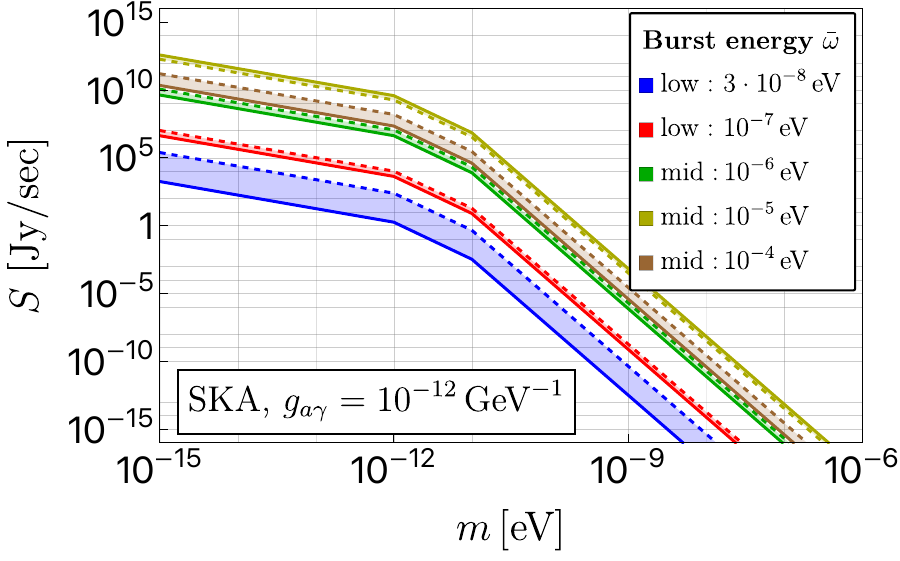}  \,
  \includegraphics[width=0.48\textwidth]{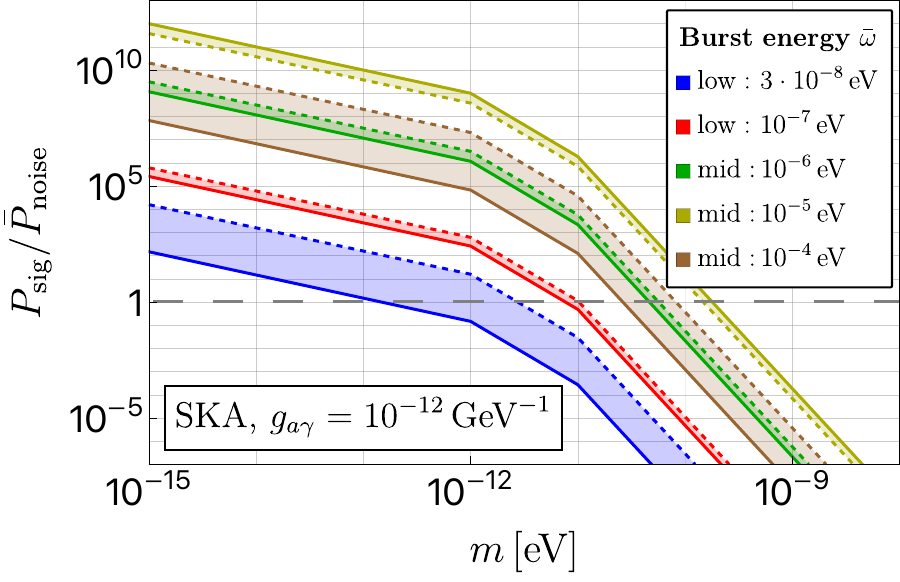}  
	\caption{(Left) Total integrated photon flux density $S$, as described by Eq.~\eqref{eq:S_flux}. (Right) Detection signal-to-noise ratio SNR = $P_{\rm sig}/\overline{P}_{\rm noise}$ (right panel) in the sensitivity range of SKA-low (blue, red) as well as SKA-mid (green, yellow, brown), as a function of axion mass $m$ for an average transient emission burst energy $\bar{\o}$. The thick (dashed) lines correspond to burst emission energy spread of $\d\o = \bar{\o}/10$ ($\d\o = \bar{\o}$). D$a$B is considered to arise from transient sources with cosmological burst rate of $R_{\rm burst} \propto (1+z)^{-1}$ and with $\mathcal{F} = 0.1$.
 }
	\label{fig:gammaflux}
\end{center}
	\vspace{-4mm}
\end{figure}

The experimental sensitivity can be conventionally estimated via by comparing the total power in the signal, $P_{\rm sig}$, to the power in the noise, $P_{\rm noise}$. The signal power can be calculated from the differential flux as
\begin{equation} \label{eq:Psignal}
    P_{\rm sig} = \eta A S~.
\end{equation}
The integrated flux density, $S$, is defined by\footnote{Conventionally, this is measured in Janskys (Jy) per second. Recall that $1$ Jy $=10^{-23}$ erg cm$^{-2} \simeq 6.2\cdot 10^{-12}$ eV cm$^{-2}$.} 
\begin{align} \label{eq:S_flux}
    S \equiv \int_{\o_1}^{\o_2} d\o 
        \o\,\frac{d\phi_\g}{d\o}~.
\end{align} 
We estimate the flux density $S$ in Fig.~\ref{fig:gammaflux} for several choices of average transient burst emission energy $\bar{\o}$.

The system's noise can be described by Dicke's radiometric receiver formula~\cite{Dicke:1946glx}
\begin{equation} \label{eq:Pnoise}
    P_{\rm noise} = T_{\rm sys} \sqrt{\frac{\D\n}{t_{\rm obs}}}~,
\end{equation}
where $\D\n$ is the frequency bandwidth in the experiment, $t_{\rm obs}$ is the observation time, $T_{\rm sys}$ is the system noise temperature. We consider the following contributions to the system noise temperature
\begin{equation}
    T_{\rm sys} = T_{\rm atm} + T_{\rm CMB} + T_{\rm bknd} + T_r~,
\end{equation}
where $T_{\rm atm} \simeq 3\,{\rm K}$~\cite{Ajello:1995aaa} is the atmospheric sky emission temperature, $T_{\rm CMB} = 2.725\,{\rm K}$ is the cosmic microwave background (CMB) radiation temperature, $T_{\rm bknd}$ is the sky brightness temperature contributed by emission of all ``background'' radio sources and $T_r$
is the noise contributed by radiometric receiver itself. For SKA-low (mid), we take an approximate averaged contribution over frequency bands of $T_r \simeq 20$ ($40$)~K~\cite{Braun:2019gdo}. The background source emission temperature can be estimated as \cite{Braun:2019gdo,Ghosh:2020hgd}
\begin{equation}
    T_{\rm bknd} \simeq 60\,{\rm K}\left(\frac{300\,{\rm MHz}}{\nu}\right)^{2.55}~.
\end{equation}

Converting the noise power $P_{\rm noise}$ in Eq.~\eqref{eq:Pnoise} to an equivalent flux, we arrive at the red lines in Fig.~\ref{fig:summary} and Fig.~\ref{fig:summary_decay}. 
As a simple noise model for our study, we use the noise averaged over the frequency range of each experiment
\begin{equation}
    \bar{P}_{\rm noise} 
        \equiv \frac{1}{\D\n}\int P_{\rm noise}(\nu)d\nu.
\end{equation}
For SKA-low (SKA-mid) we have $\bar{P}_{\rm noise} \simeq 0.3$ ($0.17$) eV/sec. 

\begin{figure} 
\begin{center}
  \includegraphics[width=0.9\textwidth]{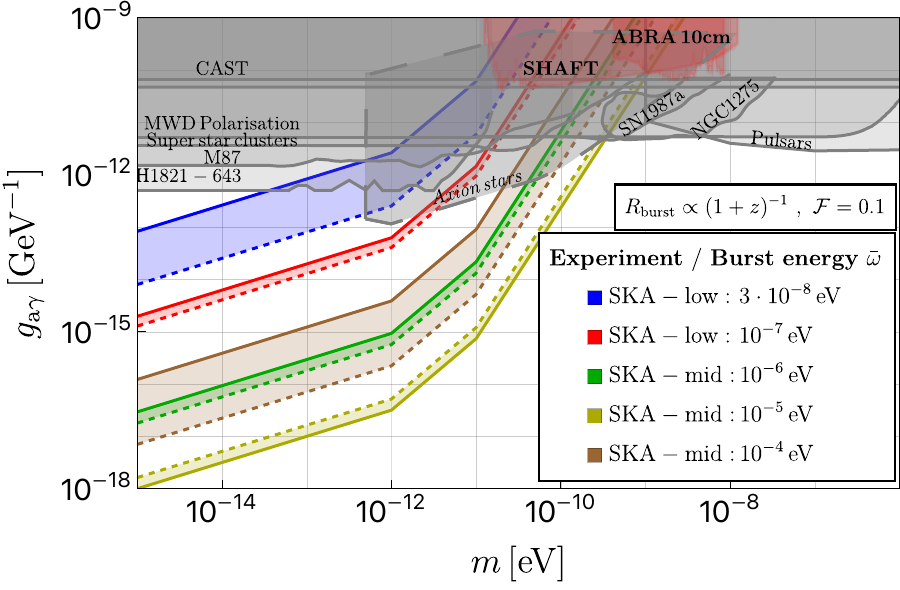}  \caption{Sensitivity estimation for SKA-low (blue, red) and SKA-mid (green, yellow, and brown) to detect D$a$B, as a function of axion mass $m$ for average transient burst energy $\bar{\o}$. The dominant axion-photon conversion process in this region is through the Galactic magnetic B-fields (see Eq.~\eqref{eq:dphido_Bfield}). The thick (dashed) lines correspond to transient burst energy spread of $\d\o = \bar{\o}/10$ ($\d\o = \bar{\o}$). D$a$B is considered to arise from transient sources with cosmological burst rate of $R_{\rm burst} \propto (1+z)^{-1}$ and with $\mathcal{F} = 0.1$. Existing constraints are overlaid in gray including astrophysical limits (only leading limits shown, see e.g. Ref.~\cite{AxionLimits,Adams:2022pbo} for a more comprehensive list) from Chandra (H1821-643~\cite{Reynes:2021bpe}, 
M87~\cite{Marsh:2017yvc}, and 
NGC1275~\cite{Reynolds:2019uqt} sources), 
super star clusters~\cite{Dessert:2020lil}, 
magnetic white dwarf (MWD) polarization~\cite{Dessert:2022yqq}, 
CAST~\cite{CAST:2007jps,CAST:2017uph}, 
pulsar polar-cap cascades~\cite{Noordhuis:2022ljw}, and SN1987a~\cite{Payez:2014xsa,Hoof:2022xbe}, direct detection limits (red) from 
SHAFT~\cite{Gramolin:2020ict} and 
ABRA 10cm~\cite{Salemi:2021gck} as well as cosmology-dependent limit from axion star explosions inducing early-Universe heating of the intergalactic medium shown with gray long-dashed line~\cite{Du:2023jxh,Escudero:2023vgv}.}
	\label{fig:sens}
\end{center}
	\vspace{-4mm}
\end{figure} 

Defining the signal-to-noise-ratio parameter ${\rm SNR}\equiv P_{\rm sig}/\bar{P}_{\rm noise}$, we also illustrate the sensitivity region for SKA for the same benchmarks in $\bar{\o}$ in the right panel of Fig.~\ref{fig:gammaflux}. 
As before, the slope of the lines changes when the behavior of the photon conversion probability changes, near $10^{-12}-10^{-11}\,{\rm eV}$, as seen in Fig.~\ref{fig:Pphigamma}(b).
In order to determine sensitivity to axions contributing to D$a$B, as before we vary coupling $g_{a\g}$ values to find sensitivity regions where SNR$>2$. The results are shown in Fig.~\ref{fig:sens}
We observe that SKA can efficiently detect D$a$B signatures for sources with central burst energies $\bar{\o}\simeq 10^{-8}-10^{-4}\,{\rm eV}$ over a wide range of axion masses $m\lesssim 10^{-9}\,{\rm eV}$.
We note that the precise sensitivity depends on the assumed cosmological population distribution of transient burst sources, as further discussed in App.~\ref{app:RateFunc}.

\section{Case study: axion star bosenovae}
\label{sec:astarbosenova}

As demonstrated, photons arising from D$a$B constitute promising targets for future searches. 
Our analysis can hence be utilized to set constraints on and probe the existence of transient sources of relativistic axion emission as well as their population distributions originating from distinct theories of new physics.

As an example, we outline the analysis steps for the scenario of relativistic axions emitted from the bosenovae that are the endpoint evolution of collapsing boson stars~\cite{Eby:2016cnq,Levkov:2016rkk}, discussed in Sec.~\ref{ssec:Bosenova}. Importantly,
the axion emission spectrum of the source in this case, $dN_a(\o)/d\o$, has been precisely determined by dedicated numerical simulations~\cite{Levkov:2016rkk}. We can model the leading relativistic peak as shown in Fig.~\ref{fig:bosenovaspectrum}, which has central axion emission energy of $\bar{\o}\simeq 2.2m$ and energy spread of $\d\o\simeq 0.4 m$, using Gaussian distribution described in Eq.~\eqref{eq:dNado_Gaus}. On the other hand, the cosmological history and transient rate of bosenovae remains poorly understood and can depend on multiple complex processes. For illustration, we adopt a power-law cosmological bosenovae source distribution of Eq.~\eqref{eq:fzmodels} with parameter $p=-1$, and analyze the resulting constraint on $\Fcal$ assuming non-observation of signals in experiments.

With above approximations, one can compute the total flux in D$a$B using Eq.~\eqref{eq:dphido_Gaus}. Assuming boson stars are formed from ultralight particles, $m\ll{\rm eV}$, and the energy scale of emission is of order $\bar{\o}\sim 2.2 m$, we will be interested in low-energy signals detectable by SKA. In this parameter range, axion decays rather than Galactic magnetic field axion conversion dominates the production of photons. Therefore, we employ Eq.~\eqref{eq:dphigdo_decay} to determine the total flux of photons, and Eq.~\eqref{eq:Psignal} as well as Eq.~\eqref{eq:S_flux} to determine the total signal power. 

\begin{figure} 
\begin{center}
  \includegraphics[width=0.9\textwidth]{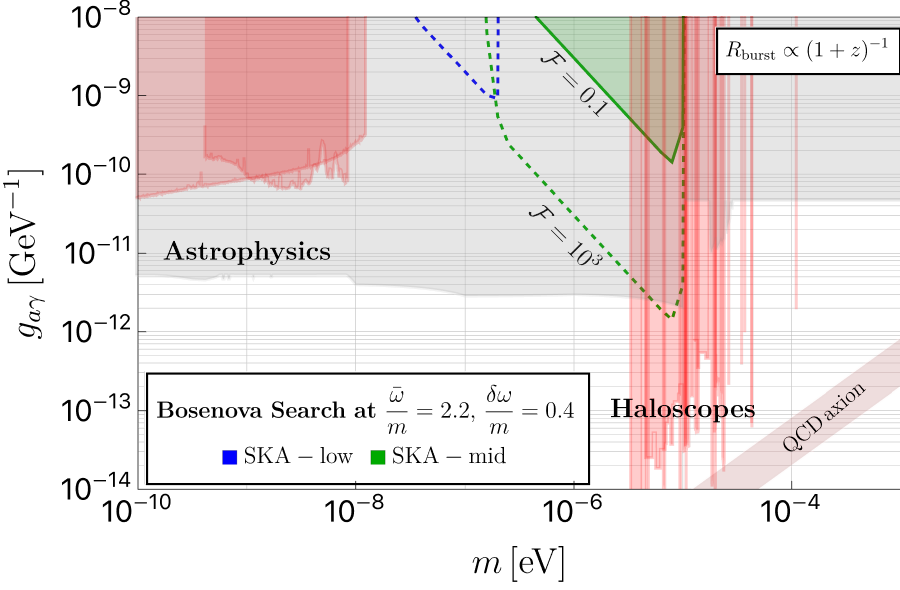} 
\caption{Sensitivity estimate for photons from the D$a$B associated with axion bosenovae, with emission spectrum modeled as Gaussian with average energy $\bar{\o}=2.2m$ and spread $\d\o=0.4 m$, considering SKA-low (blue lines) or SKA-mid (green). In this range, the flux is dominated by photons produced from axion decay (see Eq.~\eqref{eq:dphigdo_decay}). The solid lines correspond to $\Fcal=0.1$, while the dashed lines have $\Fcal=10^3$. D$a$B is considered to arise from bosenovae with cosmological burst rate of $R_{\rm burst} \propto (1+z)^{-1}$. Constraints from astrophysics and haloscopes are further described in the caption of Fig.~\ref{fig:sens2}.}
	\label{fig:Bosenova}
\end{center}
	\vspace{-4mm}
\end{figure} 

We vary $g_{a\g}$ to determine the sensitivity region where ${\rm SNR}\equiv P_{\rm sig}/\bar{P}_{\rm noise} > 2$. The results are shown in Fig.~\ref{fig:Bosenova} for SKA-low (blue) as well as SKA-mid (green). For the benchmark we used above, $\Fcal=0.1$, the best projected sensitivity in the SKA range is $g_{a\g}\gtrsim 10^{-10}\,{\rm GeV}^{-1}$ at $m\simeq 10^{-5}\,{\rm eV}$, which is weaker than existing astrophysical constraints (shaded gray region). To reach the sensitivity in axion coupling of the order of $g_{a\g}\simeq 10^{-12}\,{\rm GeV}^{-1}$, D$a$B DM fraction would need to exceed $\Fcal\simeq 10^{3}$, as can be readily seen from SNR scaling of $\propto \Fcal g_{a\g}^2$. For illustration, in Fig.~\ref{fig:Bosenova} we also display the sensitivity reach considering the unphysical benchmark of $\Fcal=10^{3}$. The results are similar for other cosmological bosenovae rate distribution functions, including Gaussian analyzed in App.~\ref{app:RateFunc}.

This example illustrates how D$a$B can be exploited to constrain the distribution of relativistic transient axion bursts from a source population with a known emission spectrum, in this case bosenovae. Such search also represents a complementary approach to existing ones that can probe couplings near the astrophysical limits. 
Based on the results above, we expect improved sensitivity for more energetic axion emission bursts, with $\bar{\o}\gg m$ (see Sec.~\ref{ssec:SKA}), or for astrophysical sources giving rise to X-ray or gamma-ray signals (see Fig.~\ref{fig:summary} and Fig.~\ref{fig:summary_decay}). 
Extrapolating this study to larger $m$, our results also motivate high-sensitivity next-generation experiments at frequencies above 15 GHz where SKA-mid loses sensitivity, which may lead to novel constrains on the D$a$B. We leave detailed investigation of such sources to future work.

%%%%%%%%%%
\section{Complementary search prospects}
\label{sec:additional}
%%%%%%%%%%

%%%
\subsection{Axion-electron and other couplings}
%%%

While we have focused on axion couplings to photons, other axion couplings (see Eq.~\eqref{eq:othercoup}) could give rise to different novel signals associated with D$a$B and call for dedicated analyses. 

As an example, consider the coupling to electrons of the form $\Lcal \supset (g_{a e}/2m_e) (\partial_\mu a) \bar{e}\g^\m\g_5 e$, where $m_e$ and $e$ are the electron mass and field, and $g_{a e}$ is the dimensionless axion-electron coupling. If the axions have mass greater than twice the electron mass, $m\gtrsim 2m_e \simeq {\rm MeV}$, this creates a new decay channel for the D$a$B and could be searched for in a wide range of experiments. Note that if the electron coupling is present at tree level, there is no analogue of magnetic-field conversion that we described for photon couplings. However, photon couplings could be induced beyond tree level, e.g. through a loop of electrons with tree-level coupling given by Eq.~\eqref{eq:othercoup}, implying a potentially interesting interplay between the two~\cite{Calore:2020tjw}.

The typical decay length for the process $a\to \bar{e}e$ is given by~\cite{Altmann:1995abc,Lucente:2021hbp} 
\begin{equation}
    \ell_e = \dfrac{\gamma v_{\rm burst}}{\Gamma_{e^+e^-}} = 8\cdot10^{-5}\,{\rm pc} \left(\dfrac{10^{-13}}{g_{a e}} \right)^2 \left(
    \dfrac{20~{\rm MeV}}{m}\right)^2
    \left(\dfrac{\o}{100~{\rm MeV}}\right)
    \left[\dfrac{1-\left(\dfrac{m}{\o}\right)^2}
    {1-\left(\dfrac{2m_e}{m}\right)^2}\right]^{1/2}\,,
\end{equation}
where we used\footnote{This also matches Ref.~\cite{Bauer:2017ris}, taking $g_{ae} = c_{\ell\ell}^{\rm eff}m_\ell/\Lambda$.} $\G_{e^+e^-}\simeq (g_{a e}^2/8\pi)\sqrt{m^2-4m_e^2}$.
Compared with the case of axion decays to photons, c.f. Eq.~\eqref{eq:decaylength}, the decay length associated with axion electron decays tends to be significantly shorter, but is restricted to sizable $m\gtrsim 2m_e$. 
The flux of electrons and positrons from axion decays is similar to the photon case in Eq.~\eqref{eq:dphigdo_decay}, and is given by
\begin{equation} \label{eq:dphiedo_decay}
    \ddx{\phi_e}{\omega}\Big|_{\rm decay} = \int_0^\infty dz 
        (1+z)\ddx{N_a(\omega(1+z))}{\omega} R_{\rm burst}(z)
                \left|\frac{dt}{dz}\right| 2P^{(e)}_{\rm decay}(R(z),\o(1+z)) ~,
\end{equation}
where $P_{\rm decay}^{(e)}(R,\o) = \left[1-\exp\left(-R/\ell_e(\o)\right)\right]$ is they decay probability.
The flux reduces, as in Eq.~\eqref{eq:dphigdo_decay_appx}, to
\begin{equation} \label{eq:dphiedo_decay_appx}
    \frac{d\phi_e}{d\o} \Big|_{\rm decay} \simeq 2P_{\rm decay}^{(e)}(R(\bar{z}),\o(1+\bar{z}))\frac{d\phi}{d\o}~,
\end{equation}
when $P_{\rm decay}^{(e)}$ is independent of redshift, e.g. for a rapid epoch of bursts at a fixed redshift $z=\bar{z}$. Here we sum the flux of electrons and positrons together.

\begin{figure}
\begin{center}
  \includegraphics[width=1\textwidth]{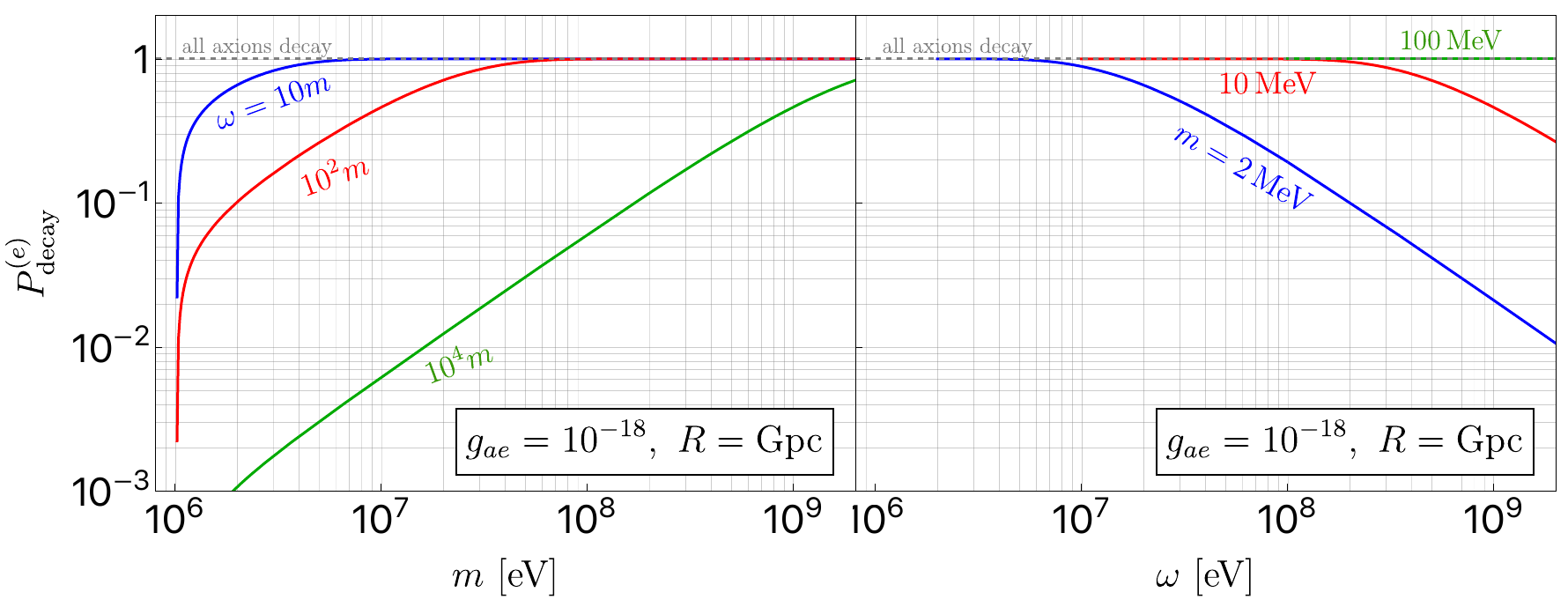}  
	\caption{Axion-electron decay probability, $P_{\rm decay}^{(e)}$, as a function of axion mass $m$ and a fixed energy $\o$ (left panel), and as a function of energy $\o$ at fixed mass $m$ (right panel). Reference values of  coupling $g_{ae}=10^{-18}$ and distance $R={\rm Gpc}$ are considered.}
	\label{fig:Pdecay_ee}
\end{center}
	\vspace{-4mm}
\end{figure} 

For cosmological distances $R\sim {\rm Gpc}$, the decay probability generally tends to unity $P_{\rm decay}^{(e)} \to 1$ when the axion-electron coupling values are near the current experimental limits for light axions, $g_{a e} \simeq 10^{-13}$ (see e.g.~\cite{Capozzi:2020cbu}). Hence, one expects nearly all sufficiently heavy axions contributing to D$a$B to decay to electrons. For illustration, in Fig.~\ref{fig:Pdecay_ee} we demonstrate that even for a significantly smaller coupling  $g_{a e}=10^{-18}$ a significant fraction of the D$a$B is expected to decay, providing opportunities for related signatures.

Electrons and positrons produced by the D$a$B can serve as novel efficient sources of photons. For astrophysical sources inside the Galaxy, primary mechanisms include interactions with
interstellar medium gas, Galactic magnetic fields, or interstellar radiation fields including the cosmic microwave background (CMB). These can lead to production of X-ray and $\g$-ray photons, serving as targets for experiments like XMM-Newton, INTEGRAL and COMPTEL, as has been shown for positrons originating in decays of astrophysical supernova axions~\cite{DelaTorreLuque:2023nhh,DelaTorreLuque:2023huu}. 
Our analysis also highlights the significance of distinct sources producing axions contributing to D$a$B and their energetics. As we have demonstrated, for example, axion decay probability can be dramatically different between the ultrarelativistic source emission with $\omega \gg m$ and semi-relativistic source emission with $\omega \sim m$.

However, the D$a$B flux need not originate from within the Galaxy, particularly in case of dark sector sources, allowing for other search approaches as we outline. In analogy to the scenario of decaying DM~(e.g.~\cite{Cirelli:2012ut,Ando:2015qda,Blanco:2018esa}), D$a$B contributions to extragalactic $\g$-ray background~can arise from Inverse Compton scattering off the CMB. Unlike decaying DM case, the origin of decaying particles is not associated with a constant DM density but depends on the cosmological redshift distribution of axion emission sources as specified by $R_{\rm burst}$ model. We can thus estimate the flux of photons of energy $\o$ in generality as
\begin{align} \label{eq:xxx}
    \ddx{\phi_\g}{\omega}\Big|_{\rm decay} 
    =&~ \int_0^\infty dz 
        (1+z)
        R_{\rm burst}(z)
                \left|\frac{dt}{dz}\right|  \\
        &\times \int d\o_e \int d\o_{\rm CMB} \int_{\o_e}^\infty d\o' P_{\rm decay}^{(e)}(R(z),\o'(1+z)) \frac{dN_a(\o'(1+z))}{d\o'} Q(\o,\o',\o_e,\o_{\rm CMB})~, \notag
\end{align}
where $Q(\o,\o',\o_e,\o_{\rm CMB})$ is a function characterizing the photon production from electron and positron conversion originating from D$a$B associated with Inverse Compton scattering against CMB photons. Here, the integrations are taken over the distribution of CMB energies $\o_{\rm CMB}$, electron energies $\o_e$, and axion energies $\o'$. The flux of photons will also experience attenuation. We leave detailed computations of realizations of such scenarios and discussion of resulting constraints for future work. 
 
Other potentially promising scenarios for observations can also arise from axions coupled to gluons or quarks, as is the case for QCD axions. In such case, axions can decay hadronically. For $m \lesssim 1$~GeV, the dominant decays are $a \rightarrow 3 \pi^0$ and $a \rightarrow \pi^+\pi^-\pi^0$~\cite{Bauer:2017ris}, with neutral pions subsequently decaying primarily into photons $\pi^0 \rightarrow \gamma\gamma$ and charged pions into muons and neutrinos. We leave detailed discussion of these possibilities for future studies.

%%%
\subsection{Direct detection}
\label{ssec:directdetection}
%%%

While we have focused on indirect D$a$B searches, direct searches are possible and complementary. Here, we briefly describe prospects and advantages of D$a$B direct detection.  

As computed in Sec.~\ref{sec:DMflux}, the total flux from D$a$B is generally sub-leading compared to the flux of local DM in the solar neighborhood vicinity. However, in some scenarios, we found that D$a$B flux can even be larger than the expected local DM flux.
A direct search for D$a$B in terrestrial axion experiments could be viable for a number of reasons. First, the frequency range of interest is shifted from around axion mass $m$, associated with non-relativistic cold axion DM described by Maxwell-Boltzmann distribution with velocity dispersion of $v_0 \simeq 220$~km/s~(e.g.~\cite{Gelmini:2015zpa} for review), to $\o\gg m$ for relativistic axions, allowing for additional sensitivity to very light fields. Further, the broad distribution of axion energies contributing to D$a$B is also distinct from the narrow distribution expected of cold DM, see Fig.~\ref{fig:DMfluxratio}. 
It is also worth noting that the very local DM density, on scales of $\sim$AU-pc within our solar neighborhood, is not well-known~\cite{Adler:2008rq,Pitjev:2013sfa}, and can deviate from the conventionally-assumed value of $\rho_{\rm loc}$. 
 
D$a$B direct detection searches are complementary to direct detection searches of relativistic axions from transient sources, as analyzed in~Ref.~\cite{Dailey:2020sxa,Eby:2021ece,Arakawa:2023gyq}. Importantly, the presence of D$a$B constitutes continuous signals with expected constant density in experiments. On the other hand, transient signals result in temporary flux density increases of axions traversing the experiments, which depending on effects such as wave-spreading can be nearly instantaneous or lasting even months or years~\cite{Eby:2021ece,Arakawa:2023gyq}. While the D$a$B flux depends on the population distribution of axion emission sources, persistent associated signals do not require existence of a particular transient source emitting axions recently or near the Earth. We also note searches for D$a$B are complementary to \emph{cosmological} sources of relativistic axions, the so-called cosmic axion background~\cite{Conlon:2013isa,Marsh:2013opc,Dror:2021nyr}, which are limited by the number of relativistic degrees of freedom in the early Universe, $N_{\rm eff}$, and also are diminished by significant redshifting of axion energies when $z\gtrsim 1000$. 

For axions of mass $m\lesssim {\rm eV}$, the integration of coherent oscillations over long timescales poses a significant source of experimental sensitivity. For such oscillations this is characterized by a coherence timescale $\tau\simeq 2\pi(m v^2)^{-1}$ for axion velocity $v$, which corresponds to an oscillator quality factor $Q\equiv v^{-2}$. For cold DM in the solar neighborhood vicinity, $Q_{\rm lDM}\simeq v_{\rm lDM}^{-2} \simeq 10^{6}$, and for transient bosenovae the effective coherence can be nearly as long due to wave-spreading effects~\cite{Eby:2021ece,Arakawa:2023gyq}. 
The D$a$B, however, is expected to be largely composed of incoherent waves from sources across the Universe, implying $Q\simeq \Ocal(1)$. This is also the case for the cosmic axion background~\cite{Conlon:2013isa,Marsh:2013opc,Dror:2021nyr}. These effects may pose challenges for direct detection, and call for further investigations. 

We further stress that the distribution of axion energies in the D$a$B can encode unique information about the cosmological distribution of sources, see Fig.~\ref{fig:DMfluxratio}. In the event of a successful cold axion DM detection, a dedicated D$a$B search carries the potential to unlock substantial novel information related to the cosmological production, SM couplings, and self-interactions of axions. We leave detailed study of these topics for future work. 

\section{Conclusions}
\label{sec:conclusions}

While conventional axion searches have primarily focused on cold axions constituting DM, relativistic axions offer novel rich opportunities to probe fundamental physics and theories. 
We have developed a general framework to describe D$a$B formed from accumulation of emitted relativistic axions from populations of distinct historic transient sources. Utilizing a small number of key input parameters that describe the cosmological population and emission properties of the sources, depending on the theory, our framework allows to easily extract the energy density and flux of the resulting axions in the vicinity of terrestrial and space-based experiments. We show that our analysis allows to comprehensively and systematically explore a wide range of D$a$B sources, include those related to astrophysics such as supernovae or NS mergers, as well as originating from dark sector dynamics such as axion star bosenova explosions.
 
For axions coupled to photons, we established sensitive constraints
on the D$a$B flux associated with light axions of mass $m\lesssim 10^{-3}\,{\rm eV}$ originating from energetic sources with $\o\gtrsim{\rm MeV}$ considering axion-photon conversion and subsequent photon detection within a range of experiments such as COMPTEL, NuSTAR, XMM-Newton, INTEGRAL, EGRET and Fermi. We demonstrate that future experiments, including SKA, JWST, XRISM, Vera C. Rubin Observatory and AMEGO/e-ASTROGAM have the capability to probe axion-photon couplings with remarkable sensitivity over multiple orders of magnitude in parameter space in axion masses and couplings for sources emitting axions with energies as low as $0.1~\mu$eV.

Focusing on relativistic axion bursts originating during or after structure formation $z\lesssim 20$, we find that the details of the cosmological burst history in this parameter range play a less significant role than the total converted fraction $\Fcal$ and the axion emission spectrum $dN_a/d\o$, which largely drive the size and detection sensitivity of signals. 
This allows to derive strong constraints from non-observation of photons from the D$a$B that largely do not significantly rely on cosmological assumptions. In the event of a detection, the spectrum of the D$a$B could provide detailed information on the cosmological evolution of axions and properties of the emission sources. 
When the axion emission bursts primarily occur within a short period at redshift $z>1$, the flux can exceed the expected background flux from local DM, providing novel prospects for both direct and indirect detection that are worthy of further exploration.

Other D$a$B searches offer additional prospects and opportunities to probe theories beyond SM. In particular, different signatures, such as novel extragalatic $\g$-ray background contributions, can appear from non-photon axion couplings, motivating further analyses.
Direct detection of D$a$B will manifest distinctly from cold axion DM and is complementary to direct searches of relativistic axions from individual transient events.

We demonstrated that a broad variety of distinct sources can contribute to D$a$B, offering rich prospects for studying new physics. Many relativistic axion emission mechanisms and resulting axion emission flux sources, as well as their cosmological population distributions and statistics remain poorly understood. 
Our work highlights the necessity and serves as an impetus calling for detailed numerical studies and simulations required to systematically describe and quantify these scenarios.

\section*{Acknowledgements}
\addcontentsline{toc}{section}{Acknowledgments}

We would like to thank Jason Arakawa, John Beacom, Oindrila Ghosh, Shunsaku Horiuchi, Marianna Safronova, and Muhammad Zaheer for discussions, as well as Pierluca Carenza and Hyungjin Kim for comments on an early draft.
The work of JE was supported by the World Premier International Research Center Initiative (WPI), MEXT, Japan and by the JSPS KAKENHI Grant Numbers 21H05451 and 21K20366, as well as by the Swedish Research Council (VR) under grants 2018-03641 and 2019-02337. VT acknowledges support by the World Premier International Research Center Initiative (WPI), MEXT, Japan and JSPS KAKENHI grant No. 23K13109.
This article is based upon work from COST Action COSMIC WISPers CA21106, supported by COST (European Cooperation in Science and Technology). 
We acknowledge the use of the AxionLimits on GitHub~\cite{AxionLimits} for compilation of existing constraints.

\appendix

\section{Results for different cosmological rate functions}
\label{app:RateFunc}

\begin{figure} 
\begin{center}
  \includegraphics[width=0.9\textwidth]{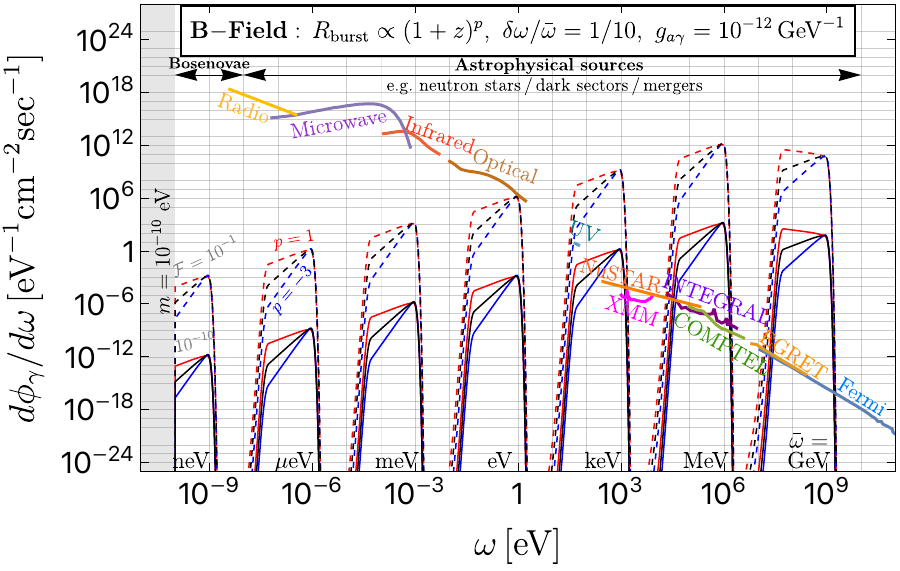}  \,\,
  \includegraphics[width=0.9\textwidth]{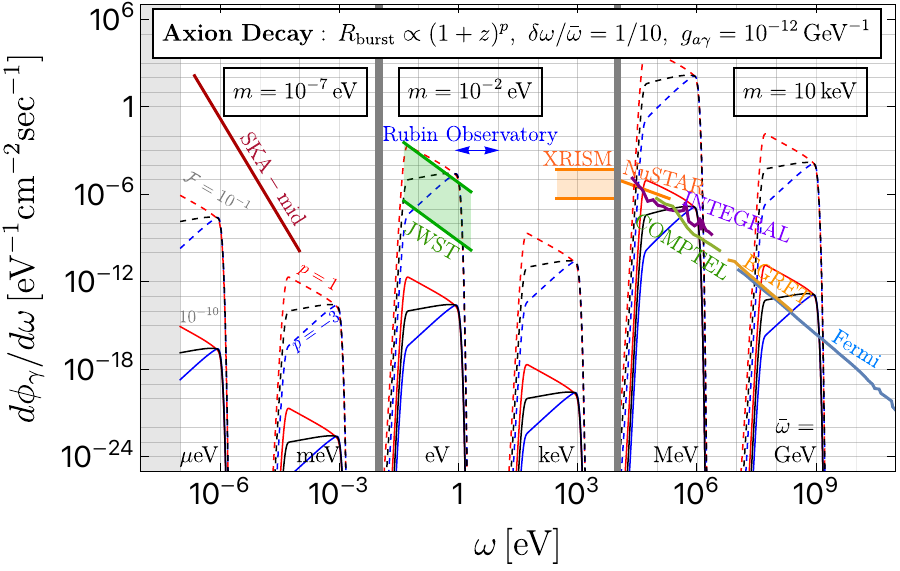} 
    \caption{ Analogous to Fig.~\ref{fig:summary} (upper) and Fig.~\ref{fig:summary_decay} (lower), but considering D$a$B arises from transient sources with cosmological burst rate of power-law with exponent $p = 1$ (red) and $p=-3$ (blue) in addition to $p=-1$ (black), $R_{\rm burst} \propto (1+z)^p$. For $B$-field conversion we take $R={\rm kpc}$, and for axion decay we assume a typical source distance of $R(z_{\rm eff})\simeq 1/H(z_{\rm eff})$ with $z_{\rm eff}$ given by Eq.~\eqref{eq:zeff} for different values of $p$.}
	\label{fig:summary_powerlaw}
\end{center}
	\vspace{-4mm}
\end{figure}

\begin{figure} 
\begin{center}
  \includegraphics[width=0.9\textwidth]{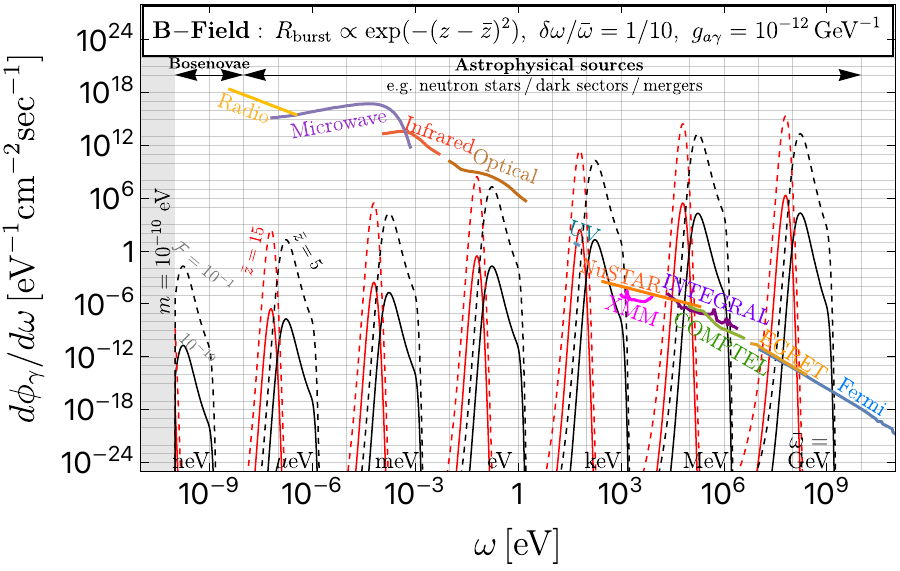}  \,\,
  \includegraphics[width=0.9\textwidth]{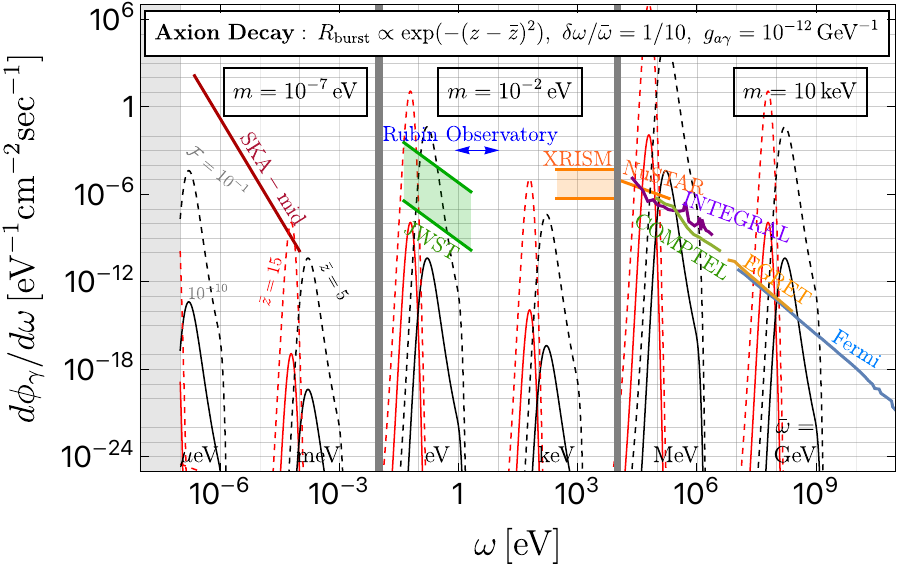} 
    \caption{Analogous to Fig.~\ref{fig:summary} (upper) and Fig.~\ref{fig:summary_decay} (lower), but considering D$a$B arises from transient sources with cosmological burst rate of gaussian with width $\d z=1$ and mean $\bar{z} = 5$ (red) and $\bar{z}=15$ (black), $R_{\rm burst} \propto \exp(-(z-\bar{z})^2/\delta z^2)$. For $B$-field conversion we take $R={\rm kpc}$, and for axion decay we assume a typical source distance of $R(z_{\rm eff})\simeq 1/H(z_{\rm eff})$ with $z_{\rm eff}$ given by Eq.~\eqref{eq:zeff}.}
	\label{fig:summary_gaussian}
\end{center}
	\vspace{-4mm}
\end{figure}

In the main text we analyzed D$a$B detection sensitivity for a power-law rate distribution (see Eq.~\eqref{eq:fzmodels}) of cosmological transient sources, considering $z_{\rm max}=20$ and power-law exponent $p=-1$. Here, we show results for different cosmological assumptions, finding similar features. In Fig.~\ref{fig:summary_powerlaw} we display the photon flux associated with magnetic field conversion as well as decays of axions from D$a$B as in Fig.~\ref{fig:summary} (upper panel) and Fig.~\ref{fig:summary_decay} (lower), but considering sources distributed according to power-law rate function with an exponent of $p=1$ (blue lines) and $p=-3$ (red). We observe that the change in exponent modifies distribution of energies, with larger $p$ implying a greater fraction of the flux lying at lower energies. Fig.~\ref{fig:summary_gaussian} illustrates the photon flux from D$a$B for sources distributed according to Gaussian rate function with $\bar{z}=5$ (red lines) and $\bar{z}=15$ (black), which is more peaked than in the power-law case.


\begin{figure} 
\begin{center}
  \includegraphics[width=0.9\textwidth]{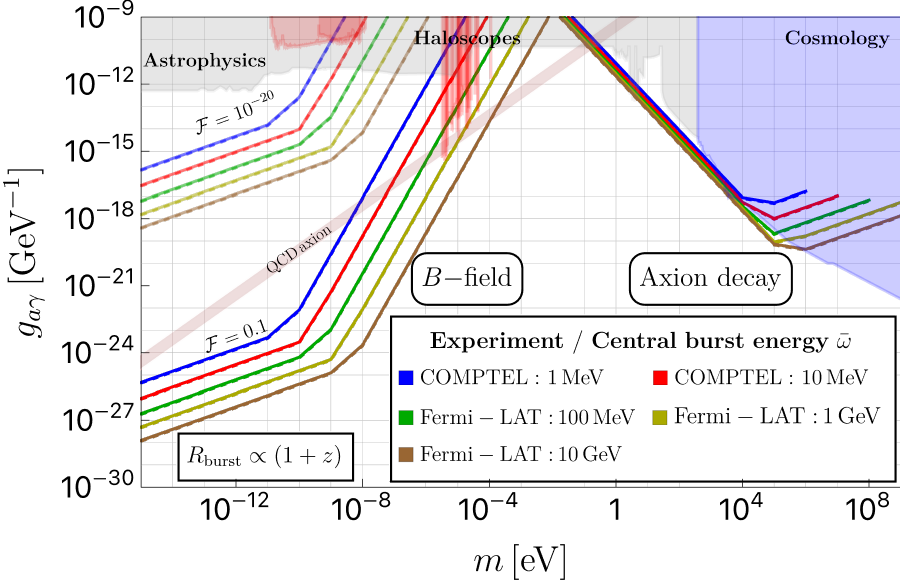}  \,\,
  \includegraphics[width=0.9\textwidth]{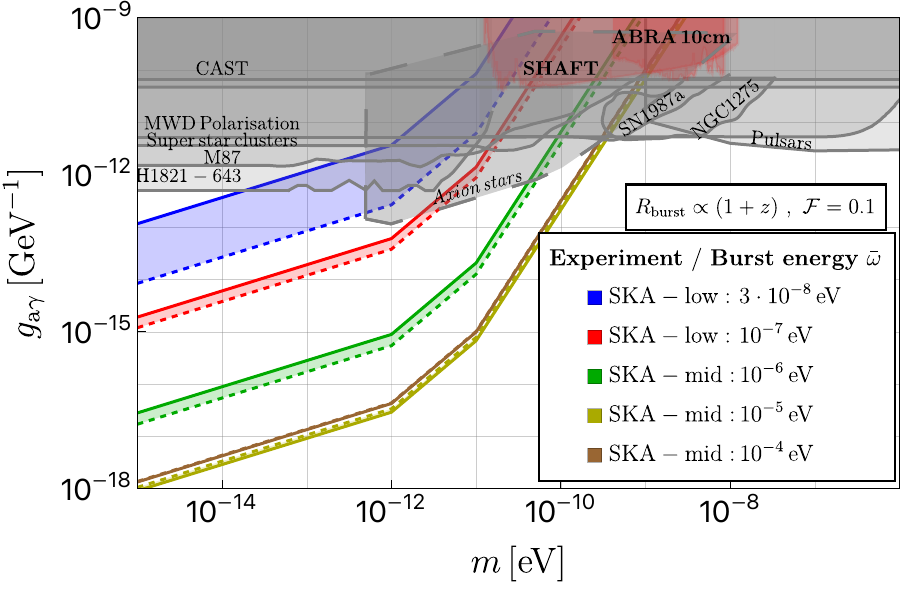} 
\caption{ Analogous to Fig.~\ref{fig:sens2} (upper) and Fig.~\ref{fig:sens} (lower), but considering D$a$B arises from transient sources with cosmological burst rate of power-law with exponent $p = 1$, $R_{\rm burst} \propto (1+z)$. For $B$-field conversion we take $R={\rm kpc}$, and for axion decay we assume a typical source distance of $R(z_{\rm eff})\simeq 1/H(z_{\rm eff})$ with $z_{\rm eff}\simeq 2/5$.}
	\label{fig:sens_a=1}
\end{center}
	\vspace{-4mm}
\end{figure}

\begin{figure} 
\begin{center}
  \includegraphics[width=0.9\textwidth]{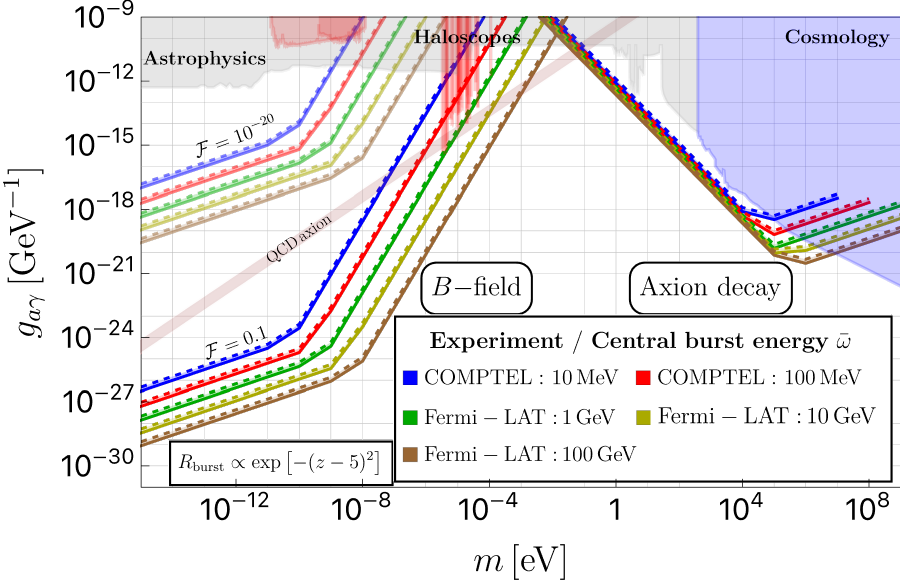}  \,\,  
  \includegraphics[width=0.9\textwidth]{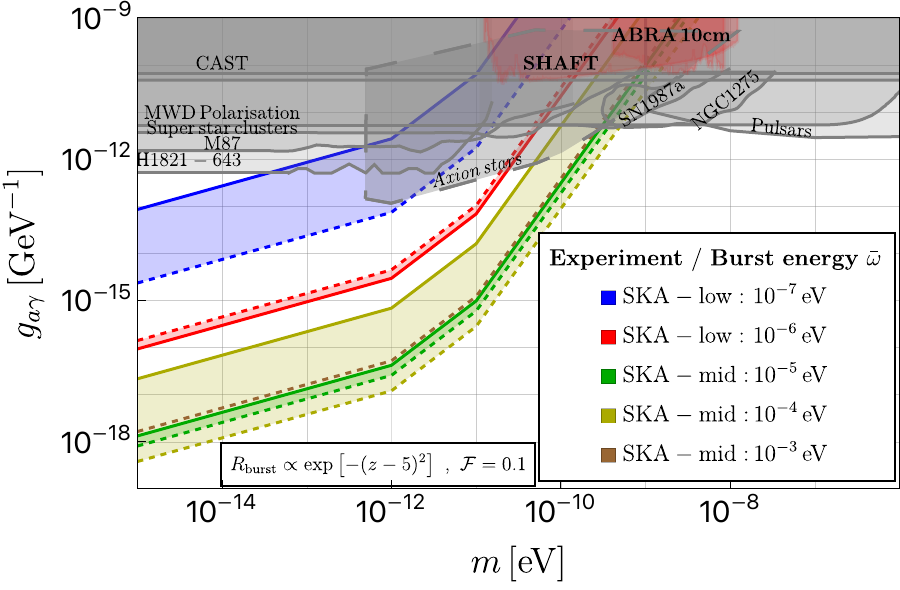}
\caption{Analogous to Fig.~\ref{fig:sens2} (upper) and Fig.~\ref{fig:sens} (lower), but using a Gaussian cosmological transient rate function $f(z)=\exp[-(z-\bar{z})^2/\d z^2]$ instead of a power law. For $B$-field conversion we take $R={\rm kpc}$, and for axion decay we assume a typical source distance of $R(z_{\rm eff})\simeq 1/H(z_{\rm eff})$ with $z_{\rm eff}$ given by Eq.~\eqref{eq:zeff}. Here, we set $\bar{z}=5$ and $\d z=1$.} 
	\label{fig:sens_zb=5}
\end{center}
	\vspace{-4mm}
\end{figure}

\begin{figure} 
\begin{center}
  \includegraphics[width=0.9\textwidth]{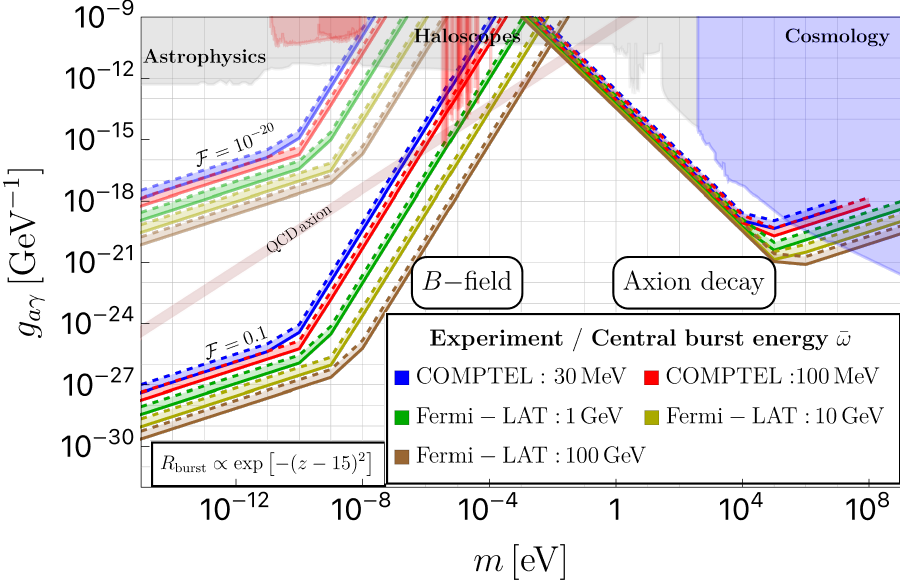}  \,\,
  \includegraphics[width=0.9\textwidth]{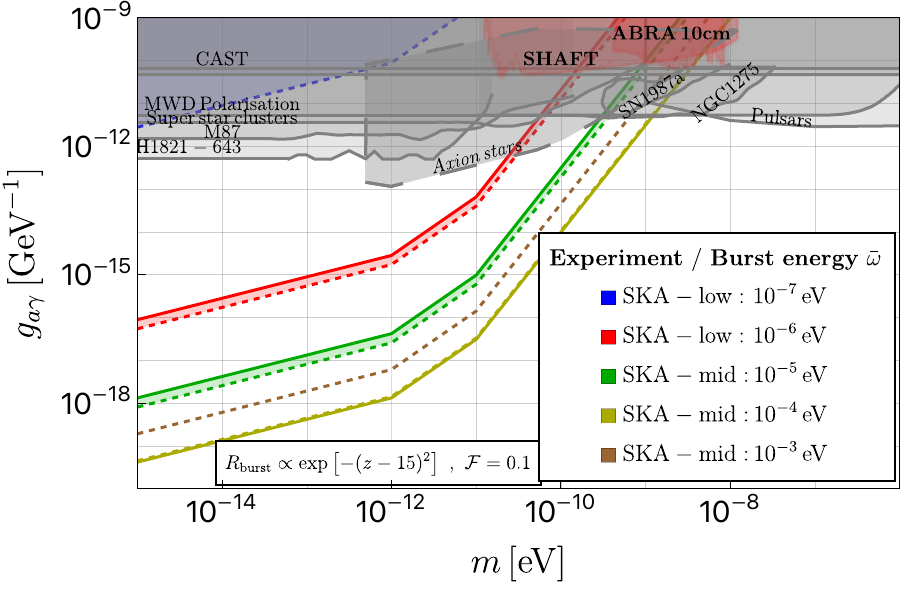} 
\caption{ Analogous to Fig.~\ref{fig:sens2} (upper) and Fig.~\ref{fig:sens} (lower), but using a Gaussian cosmological transient rate function  $f(z)=\exp[-(z-\bar{z})^2/\d z^2]$ rather than a power law. For $B$-field conversion we take $R={\rm kpc}$, and for axion decay we assume a typical source distance of $R(z_{\rm eff})\simeq 1/H(z_{\rm eff})$ with $z_{\rm eff}$ given by Eq.~\eqref{eq:zeff}. Here, we set $\bar{z}=15$ and $\d z=1$.} 
	\label{fig:sens_zb=15}
\end{center}
	\vspace{-4mm}
\end{figure}

In Fig.~\ref{fig:sens_a=1}, we display sensitivity to D$a$B as in Fig.~\ref{fig:sens2} and Fig.~\ref{fig:sens}, but considering a power-law distribution of transient sources with a different exponent parameter of $p=1$. While the sensitivity estimates are seen to be modified, our conclusions regarding the reach of present and future searches are not significantly affected. We also repeat our analysis using a Gaussian distribution of axion emission sources as described by Eq.~\eqref{eq:fzmodels}, with peak at $z=\bar{z}$ and spread $\d z$. The sensitivity estimates for this case are shown in Fig.~\ref{fig:sens_zb=5} and Fig.~\ref{fig:sens_zb=15} for $\bar{z}=5$ and $\bar{z}=15$, respectively. In both cases we consider $\d z=1$, noting that that our results do not depend sensitively on this choice.
Importantly, the peak of the energy spectrum is always located at $\bar{\o}/(1+\bar{z})$, hence the same experiments can be utilized to search for high-redshift (i.e. $\bar{z}\gg 1$) sources. The detection of a peaked signal originating at $\bar{z}$ would be optimized if the search occurs near energies of order $\bar{\o}/(1+\bar{z})$.  

\section{Photon flux estimates for future experiments}
\label{app:photonest}

Here, we briefly summarize considerations for estimating differential flux $d\phi_\g/d\o$ sensitivity of future experiments as shown in e.g.~Fig.~\ref{fig:summary_future}. These results highlight their significant potential to probe a wide range of D$a$B contributions.  

\underline{Square Kilometer Array (SKA)}~\cite{Weltman:2018zrl,Braun:2019gdo,Ghosh:2020hgd}: the signal and noise flux is estimated as described in Sec.~\ref{ssec:SKA}.

\underline{Vera C. Rubin Observatory (Rubin, formerly LSST)}~\cite{Tyson:2002mzo,Mao:2022fyx}: the expected range of sensitivity to photon energy between $(1-10)\,{\rm eV}$ is shown, without detailed sensitivity estimate.

\underline{James Webb Space Telescope (JWST)}~\cite{Bessho:2022yyu,JWST:2023abc,Janish:2023kvi,Roy:2023omw}: the expected JWST background spectrum for Galactic Center observations with polar angle $\theta_{\rm GC}=10^\circ$ and $45^\circ$ are shown in supplemental material of Ref.~\cite{Roy:2023omw}. Fig.~S3, depicts that while the two differ in detail, both lie in the range $\Bcal = (0.1-10^3){\rm MJy} = 6.2\cdot(10^{-19}-10^{-15})\,{\rm eV/cm}^2$ for wavelengths $\l=(0.6-29)\,\mu{\rm m}$, which corresponds to energy range $\o=(0.04-2)\,{\rm eV}$. Considering this estimate along with a total observing time of $t_{\rm JWST} = 10^3\,{\rm sec}$ (see Tab.~1), we calculate conversion $\Bcal \simeq t_{\rm JWST} \o^2\,d\phi_\g/d\o$ to generate results shown in Fig.~\ref{fig:summary}.

\underline{X-Ray Imaging and Spectroscopy Mission (XRISM)}~\cite{XRISMScienceTeam:2020rvx}: 
simulated spectrum for XRISM observations of the core of M87 (Virgo A) galaxy can be found in
Fig.~7 of Ref.~\cite{XRISMScienceTeam:2020rvx}, which gives $\Ccal= (0.1-10)$ counts/keV/s in the energy range $0.3-10\,{\rm keV}$. The effective collecting area in XRISM is reported to be $>160\,{\rm cm}^2$ at $1\,{\rm keV}$ and $>210\,{\rm cm}^2$ at $6\,{\rm keV}$~\cite{XRISMScienceTeam:2022exn}. For our estimates, we consider an area of $A=200\,{\rm cm}^2$. To translate the relevant quantities to units of Fig.~\ref{fig:summary_future}, we use $\Ccal \simeq A\,d\phi_\g/d\o$.

\underline{All-sky Medium Energy Gamma-ray Observatory (AMEGO)} and   \underline{A space mission for} \\ \underline{MeV-GeV gamma-ray astrophysics (e-ASTROGAM)}~\cite{e-ASTROGAM:2017pxr,DeAngelis:2021abc,AMEGO:2021abc,Engel:2022bgx}: These distinct proposals for MeV-energy gamma-ray search experiments have been estimated to have similar sensitivity, prompting us to consider them together. Estimate of the sensitivity of ASTROMEV of $\Scal \simeq (2-4)\cdot10^{-12}\,{\rm erg\,cm}^{-2}{\rm s}^{-1}$ in the photon energy range $\o = (0.1 - 300)\,{\rm MeV}$ can be found in Fig.~1 of Ref.~\cite{DeAngelis:2021abc}. We use the relation $\Scal \simeq \o^2 d\phi_\g/d\o$ to convert this into flux employed in Fig.~\ref{fig:summary_future}. A more detailed sensitivity estimate for e-ASTROGAM is given in Ref.~\cite{AMEGO:2021abc} (see e.g. their Fig.~8), which agrees with the above within approximately a factor of two. See also discussion of Ref.~\cite{Engel:2022bgx}.

\bibliography{diffusealp}
\addcontentsline{toc}{section}{Bibliography}
\bibliographystyle{JHEP}

\end{document}